\documentclass[%
 reprint,
 superscriptaddress,
 amsmath,amssymb,
 aps,
 pra,
]{revtex4-2}

\usepackage{graphicx}
\usepackage{dcolumn}
\usepackage{bm}
\usepackage{hyperref}
\usepackage{color}
\usepackage{hyperref}
\hypersetup{colorlinks, linkcolor=blue}

\begin{document}

\title{Numerical investigation of a family of solitary-wave solutions for the nonlinear Schr\"odinger equation perturbed by third-, and fourth-order dispersion} 

\author{Oliver Melchert}
\email{melchert@iqo.uni-hannover.de}

\author{Ayhan Demircan}

\affiliation{
Leibniz Universität Hannover, Institute of Quantum Optics, Welfengarten 1, 30167 Hannover, Germany
}
\affiliation{
Leibniz Universität Hannover, Cluster of Excellence PhoenixD, Welfengarten 1A, 30167 Hannover, Germany
}

\date{\today}

\begin{abstract}

We study solitary wave solutions for the nonlinear Schrödinger equation perturbed by the effects of third-, and fourth-order dispersion, maintaining a wavenumber gap between the solitary waves and the propagation constant.
We numerically construct members of a family of such solitary waves, including Kruglov and Harvey's exact solution, using the spectral renormalization method and establish empirical relations between the pulse parameters.
%
%
%
%
A deeper insight into the properties of solitary waves and solitons can be obtained through collisions.
Therefore we perform pulse propagation simulations demonstrating different collision regimes. Depending on the pulses initial phase difference, this can lead to the formation of short-lived two-pulse bound states. 
While these collisions are generally inelastic, singular phase values exist at which they are elastic. 
%
%
Finally, we detail the properties of Kruglov and Harvey's soliton solution under loss, verifying earlier predictions of perturbation theory and suggesting a convergence to the soliton solution of the standard nonlinear Schr\"odinger equation in the limit of large propagation distances. 
\end{abstract}


\maketitle


\section{Introduction}

As introduced by Zabusky and Kruskal \cite{Zabusky:PRL:1965}, the term ``solitary wave'' (SW) describes pulse-like solutions of certain nonlinear wave equations which travel with constant shape and speed. In a SW, the effects of linear dispersion on the pulse envelope are balanced by the effects imposed by the nonlinearity. Quoting Scott in Ref.~\cite{Scott:CC:1979}, ``There is just enough \emph{yin} for the \emph{yang}; it is a dynamically self-sufficient object, a `thing`''.
Commonly, if two SW's are initialized so as to engange in a collision, the nonlinear interaction upon collision destroys their integrity and identity.
However, as demonstrated by Zabusky and Kruskal in terms of numerical simulations for the Korteweg-de Vries equation \cite{Zabusky:PRL:1965}, special SW's exist, coined `solitons`, that emerge from collisions with unchanged shape and speed \cite{Scott:IEEE:1973}.
The only effect of their mutual interaction is a phase-shift that both pulses acquire during their collision \cite{Zakharov:JETP:1972,Aossey:PRA:1992}.
It is this preservation of shape and speed that renders the soliton interesting from a point of view of applied science \cite{Kivshar:BOOK:2003,Mitschke:BOOK:2016}.
Solitons, and, more generally, SW's arise in diverse fields of physics such as, e.g.,
hydrodynamics \cite{Russell:BAR:1844},
plasma dynamics \cite{Zabusky:PRL:1965},
domain wall dynamics \cite{Jin:PRB:2024},
and nonlinear optics \cite{Mitschke:BOOK:2016}.

For the nonlinear equations that govern several of the above-mentionen fields, mathematical techniques have been developed that yield closed form solutions in terms of inverse problems \cite{Gardner:PRL:1967,Lax:CPAM:1968,Ablowitz:PRL:1973,Ablowitz:SAM:1974}. 
For instance, for the nonlinear Schrödinger equation (NSE), 
which, as considered below, 
models the combined effects of group-velocity dispersion (GVD) and third-order nonlinearity in
nonlinear optics \cite{Mitschke:BOOK:2016}, soliton solutions can be obtained by the inverse scattering transform \cite{Zakharov:JETP:1972,Satsuma:PTP:1974,Kaup:JMP:1975}.
While the standard NSE is fully integrable \cite{Zakharov:JETP:1972}, exhibiting an infinite number of conserved quantities, perturbation by further linear and/or nonlinear contributions destroys its exact integrability.  The dynamics of solitons in such perturbed, yet nearly integrable systems is reviewed in Ref.~\cite{Kivshar:RMP:1989}.
%
As pointed out above, collisions of true solitons in the standard NSE proceed elastically with no radiation losses, and affect only the carrier phase and peak location of both pulses \cite{Zabusky:PRL:1965,Gordon:OL:1983}. 
In contrast, SW collisions in perturbed variants of the NSE proceed inelastically  \cite{Kivshar:RMP:1989,Frauenkron:PRE:1996,Jakubowski:PRE:1997,Anastassiou:PRL:1999,Yang:PRL:2000,Dmitriev:Chaos:2002,Dmitriev:PRE:2002,Feigenbaum:OE:2004,Feigenbaum:JOSAB:2005,Dingwall:NJP:2018,Rao:PRE:2020},
demonstrating several non-integrable characteristics:
colliding SWs can exchange energy and momentum, they may suffer radiation losses, and the energy exchange between two SWs can lead to the formation of an unstable, short-lived bound state. 
%
%
%
The resulting scattering dynamics, which are very sensitive to the initial conditions, have been carefully analyzed and explained, e.g., for the perturbed NSE \cite{Dmitriev:Chaos:2002,Dmitriev:PRE:2002}, coupled NSEs \cite{Yang:PRL:2000}, Bose-Einstein condensates \cite{Edmonds:NJP:2017,Dingwall:NJP:2018}, and a general ordinary differential equation model \cite{Goodman:PRL:2007}.

Let us note that propagation constants that account for higher orders of dispersion, with negative fourth-order dispersion (4OD), facilitate a variety of additional effects such as spectral tunneling \cite{Tsoy:PRA:2007,Serkin:EL:1993}, two-frequency soliton molecules and meta-atoms \cite{Melchert:PRL:2019,Melchert:OL:2021,Melchert:OPTIK:2023}, and can even support further special types of SWs \cite{Karlsson:OC:1994,Kruglov:PRA:2018,Tam:PRA:2020,Lourdesamy:NP:2021}. Such systems are still essentially unexplored.
%
In this regard, perturbing the standard anomalous GVD NSE by a term that accounts for third-order dispersion (3OD), SW-like pulses, located within the domain of anomalous dispersion, 
suffer radiation losses to a phase-matched frequency
beyond the zero-dispersion point in the domain of normal dispersion \cite{Wai:PRA:1990,Wai:OL:1986,Akhmediev:PRA:1995,Yulin:OL:2004,Skryabin:PRE:2005}.
Considering fourth-order dispersion (4OD), it is necessary to distinguish: given anomalous GVD, positive 4OD causes SW-like pulses to radiate \cite{Akhmediev:PRA:1995}, while negative 4OD does not impede  SW propagation \cite{Karlsson:OC:1994,Akhmediev:OC:1994}.
%
It is in this latter setting, i.e.\ the NSE perturbed by a negative 4OD term, wherein which Karlsson and H\"o\"ok showed the existence of a family of SW solutions with fixed parameters and $\rm{sech}^2$-shaped envelope \cite{Karlsson:OC:1994}. 
For large values of the soliton propagation constant, these solutions were found to exhibit radiationless oscillating tails \cite{Akhmediev:OC:1994}. 
While two-soliton and multisoliton bound states were shown to exist in the parameter range allowing for oscillating tails, all bound states turned out to be unstable \cite{Buryak:PRE:1995}. 
Pulse propagation studies further suggested this SW solution to persist for small perturbations by 3OD \cite{Piche:OL:1996}.
%
More recently, Kruglov and Harvey presented an exact stationary $\mathrm{sech}^2$-shaped SW solution for the standard anomalous GVD NSE perturbed by 3OD and negative 4OD \cite{Kruglov:PRA:2018}, herafter referred to as the KH SW solution,
obtained via reduction to an ordinary differential equation for which solutions can be expressed in terms of elliptic Jacobi functions \cite{Kruglov:OC:2020}. This SW solution exhibits no nontrivial free parameter, has a fixed velocity that depends on all orders of dispersion, and reduces to the above Karlsson-H\"o\"ok soulution for vanishing 3OD. 
Under additional perturbation by self-steepening, the existence and stability of quartic and dipol-solitons was later demonstrated \cite{Kruglov:PRA:2020,Triki:PRE:2020}. 
Also, multi-quartic and multi-dipole solitons were reported in absence of self-steepening \cite{Kruglov:CSF:2023}, and the existence of periodic solutions and SWs in the NSE with 3OD, 4OD, self-steepening, and cubic-quintic nonlinearity was shown \cite{Kruglov:CSF:2022}.

%
Here, we perform numerical simulations to study the propagation dynamics of Kruglov and Harvey's $\mathrm{sech}^2$-shaped SW solution and to complement previous theoretical results \cite{Kruglov:PRA:2018}.
The article is organized as follows. 
In Sect.~\ref{sec:model} we detail the propagation model and algorithms used to perform the numerical simulations reported in Sect.~\ref{sec:results}. 
In Sect.~\ref{sec:res01} we analyzing an auxiliary linear problem, previously  considered to classify localized solutions of the NSE perturbed by 4OD in terms of the decay along their tails \cite{Akhmediev:OC:1994,Buryak:PRE:1995,Tam:OL:2019}.
We further construct members of a family of SWs, including the KH SW solution, in terms of the spectral renormalization method (SRM) \cite{Ablowitz:OL:2005}, and assess how the pulse parameters depend on their wavenmuber.
In Sect.~\ref{sec:res02} we study the interaction dynamics of the KH SW and an independent SW with different properties, obtained using the spectral renormalization method SRM. 
In Sect.~\ref{sec:res03} we consider the effect of absorption on the propagation dynamics of the KH SW and compare the
results of our simulations to those of a perturbation analysis presented earlier in Ref.~\cite{Kruglov:PRA:2018}. In addition, going beyond the predictions of perturbation theory, we here also clarify the behavior for long propagation distances in terms of a simple model.
Section~\ref{sec:summary} concludes with a summary.

\section{Model and methods \label{sec:model}}

\paragraph*{Propagation model.}
Below we consider a higher-order NSE (HONSE) for the pulse envelope $\psi\equiv \psi(z,\tau)$, including second-, third-, and fourth-order dispersion in the form
\begin{align}
i \partial_z \psi = -i\frac{\mu}{2}\psi + \frac{\beta_2}{2} \partial_\tau^2 \psi + i\frac{\beta_3}{6}\partial_\tau^3 \psi - \frac{\beta_4}{24} \partial_\tau^4 \psi - \gamma |\psi|^2 \psi, \label{eq:HONSE}
\end{align}
where $z$ is the propagation coordinate, $\tau=t-\beta_1 z$ is a retarded time, $\beta_2$, $\beta_3$, and $\beta_4$ are the parameters specifying GVD, 3OD, and 4OD, respectively. $\mu$ describes absorption ($\mu>0$), and $\gamma$ is a scalar nonlinear parameter. 
As shown by Kruglov and Harvey in Ref.~\cite{Kruglov:PRA:2018}, for $\mu=0$ and when the dispersion coefficients satisfy the conditions 
\begin{align}
    \beta_2 < 0, \quad \beta_4<0, 
    \quad \text{and,}\quad 2\beta_2 \beta_4 > \beta_3^2, \label{eq:KH_conditions}
\end{align}
%
Eq.~(\ref{eq:HONSE}) exhibits the exact SW solution
\begin{align}
    \psi_{\rm{KH}}(z,\tau) = u\,\mathrm{sech}^2\left[w (\tau-\eta-\tfrac{z}{v})\right]\,e^{i(\kappa z - \delta \tau + \phi)}, \label{eq:KH_SW}
\end{align}
with initial pulse peak-position $\eta$ and initial phase $\phi$.
%
The amplitude $u$, inverse temporal width $w$, velocity $v$ in the retarded frame, frequency offset $\delta$ and wavenumber $\kappa$ are given by \cite{Kruglov:PRA:2018}
\begin{subequations}\label{eq:KH_SOL_pars}
\begin{align}
    u &= \sqrt{\frac{9}{20 \gamma |\beta_4|}}\left( \frac{\beta_3^2 - 2 \beta_2 \beta_4}{ \beta_4}\right),\label{eq:KH_SOl_u0}\\  
    w &= \sqrt{\frac{6 \beta_2 \beta_4 - 3\beta_3^2}{10 \beta_4^2}},\label{eq:KH_SOl_w}\\  
    v &= \frac{3 \beta_4^2}{\beta_3 (\beta_3^2 - 3 \beta_2 \beta_4)},\label{eq:KH_SOl_v}\\  
    \delta &= - \frac{\beta_3}{\beta_4},\label{eq:KH_SOl_delta}\\  
    \kappa &= - \frac{6 \left( \beta_3^2 - 2\beta_2\beta_4\right)^2}{25 \beta_4^3}  - \frac{\beta_3^2 \left( \beta_3^2 - 4 \beta_2\beta_4\right)}{8 \beta_4^3}.\label{eq:KH_SOl_kappa}
\end{align}
\end{subequations}
%
As pointed out in Ref.~\cite{Kruglov:PRA:2018}, Eq.~(\ref{eq:KH_SW}) has no nontrivial free parameter, hence the parameters Eqs.~(\ref{eq:KH_SOL_pars}) are fixed by setting $\gamma$, as well as $\beta_2$, $\beta_3$, and $\beta_4$ in correspondence with conditions~(\ref{eq:KH_conditions}).
Subsequently we consider $\mu=0$, $\gamma=1~\mathrm{W^{-1}/km}$, $\beta_2=-1~\mathrm{ps^2/km}$, $\beta_3=0.5~\mathrm{ps^3/km}$, and $\beta_4=-1~\mathrm{ps^4/km}$.
For this choice of parameters, Eqs.~(\ref{eq:KH_SOl_u0})-(\ref{eq:KH_SOl_kappa}) yield
$u\approx 1.174~\mathrm{W^{1/2}}$, $w\approx 0.725~\mathrm{ps^{-1}}$ ($w^{-1}\approx1.380~\mathrm{ps}$), $v\approx -2.182~\mathrm{km/ps}$, $\delta=0.5~\mathrm{ps^{-1}}$, and $\kappa\approx 0.618~\mathrm{km^{-1}}$.
Energy and momentum, given by
\begin{align}
E(z)&=\int |\psi|^2~{\mathrm{d}}\tau,\quad\text{and},\label{eq:energy}\\
M(z)&=\frac{i}{2}\int \left( \psi \frac{\partial \psi^*}{\partial \tau} - \psi^* \frac{\partial \psi}{\partial \tau} \right)~{\mathrm{d}}\tau,  \label{eq:momentum}
\end{align}
for $\psi=\psi_{\rm{KH}}$ read $E_{\rm{KH}}\approx 2.536~\mathrm{W\,ps}$ and $M_{\rm{KH}}\approx -1.268~\mathrm{W}$. (Let us note that $M_{\rm{KH}} = -E_{\rm{KH}} \,\delta$ \cite{Kruglov:PRA:2020}.)

%
%
%
%
%

\begin{figure}[t!]
\includegraphics[width=\linewidth]{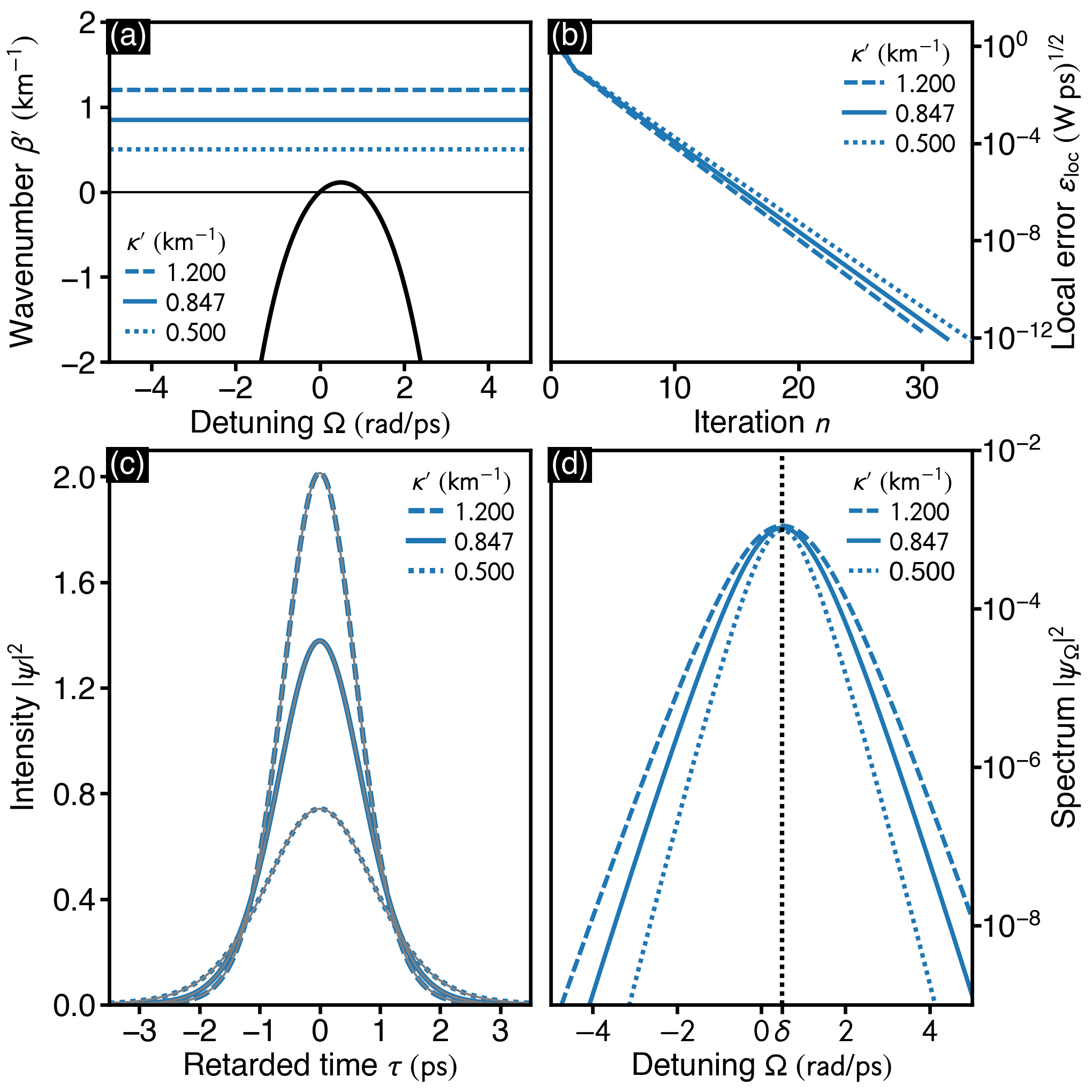}
\caption{SW solutions for Eq.~(\ref{eq:HONSE}) with $\mu=0$, $\gamma=1~\mathrm{W^{-1}/km}$, $\beta_2=-1~\mathrm{ps^2/km}$, $\beta_3=0.5~\mathrm{ps^3/km}$, and $\beta_4=-1~\mathrm{ps^4/km}$, obtained using the SRM.
(a) Modified propagation constant $\beta^\prime$, defining the reference frame in which the sought-for solitary waves are stationary.
(b-d) SRM results for selected wavenumbers using a Gaussian trial-function $\psi_0(\tau)=\exp(-\tau^2)$.
(b) Decrease of the local error $\epsilon_{\rm{loc}}$ upon iteration.
(c) Intensity of the solutions. Thin grey lines indicate fits to the three-paramter model detailed in the text.
(d) Spectrum of the solutions. 
Results for $\kappa^{\prime}=0.847~\mathrm{km^{-1}}$ produce the exact KH SW solution Eq.~(\ref{eq:KH_SW}).
}
\label{fig:01}
\end{figure}

\paragraph*{Spectral renormalization method.}
In order to obtain further SW solutions, different from the special solution $\psi_{\mathrm{KH}}$, we employ the spectral renormalization method (SRM) \cite{Ablowitz:OL:2005}, amended to work for Eq.~(\ref{eq:HONSE}) using the above parameters. Independent SW solutions for specified velocity and wavenumber will allow us to study the collision dynamics involving KH SWs in Sect.~\ref{sec:res02} below.
The SRM constitutes an iterative optimization procedure, based on Petviashvili's method for the numerical approximation of stationary solutions of nonlinear wave equations \cite{Petviashvili:SJPP:1976,Pelinovsky:PRE:2000}, which has been adapted to numerous variants of the NSE \cite{Musslimani:JOSAB:2004,Fibich:PD:2006,Amiranashvili:PRA:2013}.
%
%
%
Here, we opt for the SRM since it allows to conveniently set the wavenumber for which a SW solution is sought for.
To demonstrate the applicability of the SRM to the present setting, we first aim to reproduce the exact KH SW solution Eq.~(\ref{eq:KH_SW}).
Let us note that, when employing the SRM, it is most convenient to look for the SW in a reference frame where its location remains stationary in time. This can be achieved by considering, instead of the propagation constant $\beta(\Omega)=\tfrac{\beta_2}{2}\Omega^2 + \tfrac{\beta_3}{6}\Omega^3 + \tfrac{\beta_4}{24}\Omega^4$
of Eq.~(\ref{eq:HONSE}), the modified propagation constant $\beta^\prime(\Omega) = \beta(\Omega) - \Omega/v$ [Fig.~\ref{fig:01}(a)],
and by accounting for the corresponding shift in wavenumber by $\kappa^\prime=\kappa - \delta/v \approx 0.847~\mathrm{km^{-1}}$ \cite{Kruglov:PRA:2018}.
From the modified propagation constant shown in Fig.~\ref{fig:01}(a) we can expect to find localized solutions only above a certain threshold given by $\kappa_0\equiv \max_\Omega(\beta^\prime) = 0.112~\mathrm{km^{-1}}$.
For values $\kappa>\kappa_0$, a finite wavenumber gap separates these localized solutions from $\beta^\prime$.
We find that at $\kappa^{\prime}=0.847~\mathrm{km^{-1}}$, a trial-function of the form $\psi_0(\tau)=\exp(-\tau^2)$ rapidly converges to the exact solution $\psi_{\mathrm{KH}}(0,\tau)$ [Figs.~\ref{fig:01}(c,d)].
This is evident from the local error $\epsilon_{{\rm{loc}},n} \equiv \left( \int |\psi_{n}-\psi_{n-1}|^2~d\tau \right)^{1/2}$, which exhibits an exponential decrease with increasing iteration step $n$ of 
the SRM procedure and converges after 32 iteration steps [Fig.~\ref{fig:01}(b)].
Let us add that the global error with respect to the exact solution, i.e.\ $\epsilon_{{\rm{glob}},n}\equiv \left( \int | \psi_{\rm{KH}} - \psi_n|^2~d\tau \right)^{1/2}$, decreases in the same manner, resulting in $\epsilon_{{\rm{glob}},n=30}\approx 6\times10^{-13}~\mathrm{(W\,ps)^{1/2}}$.
Results for two further SW wavenumbers $\kappa^{\prime}=0.5~\mathrm{km^{-1}}$ and $1.2~\mathrm{km^{-1}}$ are included with Fig.~\ref{fig:01}.

\paragraph*{Propagation algorithm.}
For the pulse propagation simulations in Sects.~\ref{sec:res02}, i.e.\ when solving the initial-value problem for Eq.~(\ref{eq:HONSE}) with $\mu=0$, we employ  the conservation quantity error (CQE) method \cite{Heidt:JLT:2009,Melchert:CPC:2022}, which uses a conservation law of the considered HONSE to control adaption of the stepsize $h$. 
Specifically, we use the relative energy error 
$\delta_{\rm{E}}(z) = |E(z+h) - E(z)|/E(z)$,
wherein $E$ specifies the pulse energy Eq.~(\ref{eq:energy}) conserved by Eq.~(\ref{eq:HONSE}) in case of $\mu=0$ \cite{Kruglov:PRA:2018}.
Here, the CQE method is set to maintain the relative error $\delta_{\rm{E}}$ within the goal error range $(10^{-11}, 10^{-10})$, by decreasing the stepsize $h$ when necessary while increasing $h$ when possible.
To advance the field from position $z$ to $z+h$, we use the forth-order Runge-Kutta in the interaction picture (RK4IP) method \cite{Hult:JLT:2007}.
In Sect.~\ref{sec:res03} we instead employ the RK4IP with fixed stepsize.


\section{Results\label{sec:results}}

Subsequently, in Sect.~\ref{sec:res01}, we discuss the asymptotic behavior of the low-intensity tails of solutions to Eq.~(\ref{eq:HONSE}) in terms of a linear auxiliary equation and assess how the properties of the nonlinear localized states depend on their wavenumber. We then report results of pulse propagation simulations that clarify the collision dynamics involving KH SWs in Sect.~\ref{sec:res02}. Finally, we discuss the decay 
of the KH SW under absorption in Sect.~\ref{sec:res03}.

\subsection{SW solutions parameterized by wavenumber \label{sec:res01}}

\begin{figure}[t!]
\includegraphics[width=\linewidth]{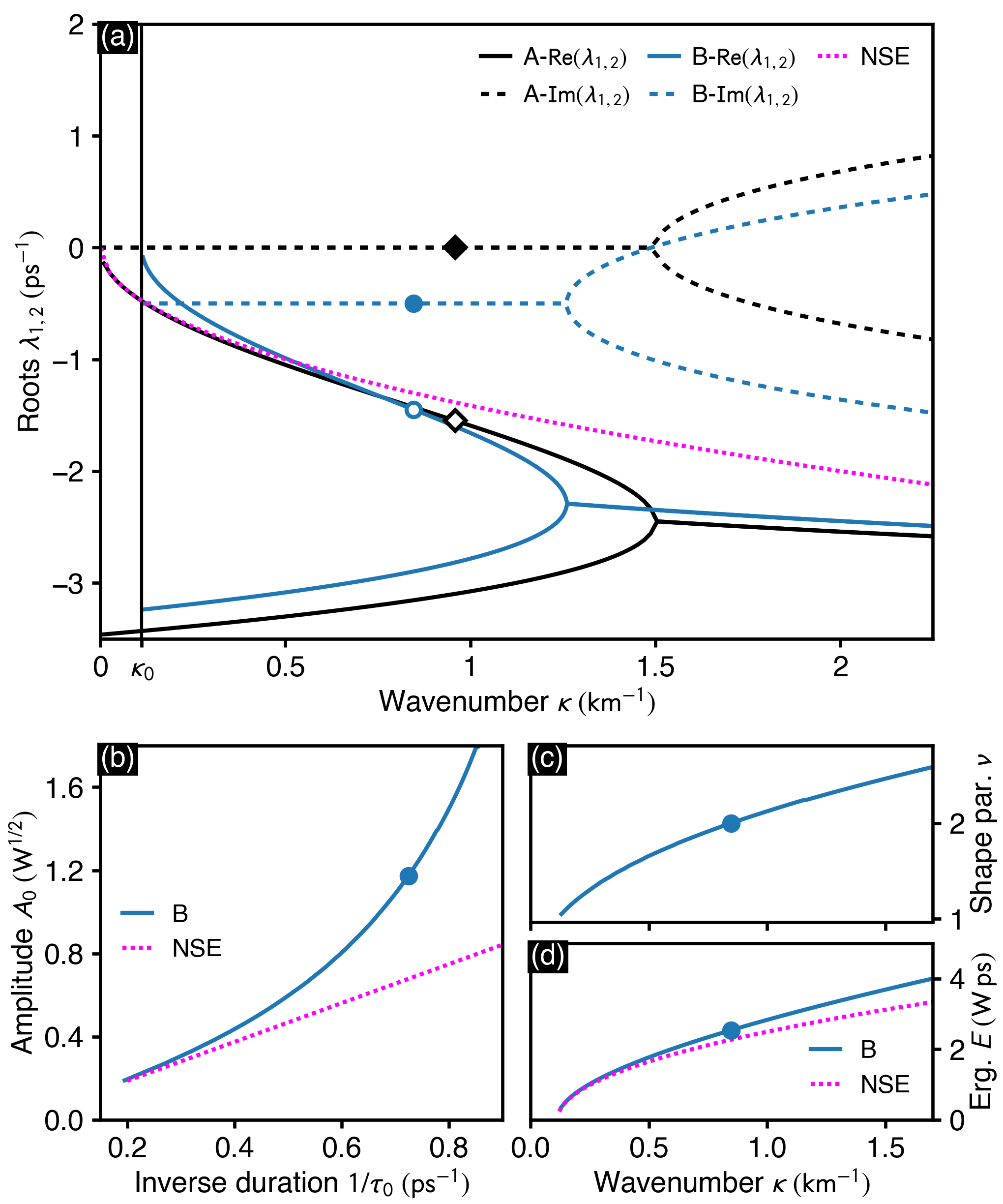}
\caption{Characteristics of localized solutions for Eq.~(\ref{eq:HONSE}) with parameters listed in Tab.~\ref{tab:table01}.
(a) Roots $\lambda_{1,2}(\kappa)$, determining the asymptotic decay according to Eq.~(\ref{eq:lin_eq}). 
Parameters corresponding to labels A and B are listed in Tab.~\ref{tab:table01}.
%
%
In case of B, nontrivial solutions exist only for $\kappa > \kappa_0=0.112~\mathrm{km^{-1}}$.
Short-dashed line (labeled NSE) indicates the limiting case of the standard NSE. 
%
Filled and open circles at $\kappa = 0.847~\mathrm{km^{-1}}$ indicate $-\delta=-0.5~\mathrm{ps^{-1}}$ and $-2w \approx -1.450~\mathrm{ps^{-1}}$, respectively. 
(b-d) Results for parameter setting B obtained using the SRM.
(b) Dependence of pulse amplitude on inverse pulse duration (labeled B).
(c) Variation of the shape parameter $\nu$ as function of wavenumber.
(d) Variation of the pulse energy as function of wavenumber (labeled B).
In (b,d) dashed lines (labeled NSE) indicate the corresponding relations for the standard NSE (see text for details).
}
\label{fig:02}
\end{figure}

\paragraph*{Asymptotic decay of the solutions.}
Neglecting the nonlinear contribution to Eq.~(\ref{eq:HONSE}), and focusing on the asymptotic decay of the localized states towards $\tau\to \infty$ in terms of an Ansatz of the form $\psi(\tau,z) \propto e^{\lambda \tau + i\kappa z}$, yields the algebraic equation 
\begin{align}
\kappa = - i \beta_1\lambda - \frac{\beta_2}{2} \lambda^2 - i\frac{\beta_3}{6} \lambda^3 + \frac{\beta_4}{24}\lambda^4. \label{eq:lin_eq}
 \end{align}
As discussed for the NSE with added 4OD \cite{Akhmediev:OC:1994,Buryak:PRE:1995,Tam:PRA:2020}, and similarly for the generalized Lugiato-Lefever equation \cite{Melchert:OL:2020,Rivas:PRA:2014}, an analysis of the four roots $\lambda_n$, with $n=1\ldots 4$, of Eq.~(\ref{eq:lin_eq}) allows to classify the localized solutions of the full nonlinear model according to the behavior of their low-intensity tails.
As pointed out in the context of the Karlsson-H\"o\"ok SW solution to Eq.~(\ref{eq:HONSE}) for $\beta_1=0$ and $\beta_3=0$, decaying solutions are characterized by roots of Eq.~(\ref{eq:lin_eq}) exhibiting a negative real part \cite{Akhmediev:OC:1994}. 
These are two out of the four roots, which we below refer to as $\lambda_1$ and $\lambda_2$, ordered so as to satisfy $\mathrm{Re}[\lambda_1]\leq \mathrm{Re}[\lambda_2]$.
However, the existence of decaying solutions of Eq.~(\ref{eq:lin_eq}) does not automatically guarantee the existence of SW-like solutions of the full nonlinear Eq.~(\ref{eq:HONSE}) \cite{Akhmediev:OC:1994}. 
While closed form solutions for the biquadratic version of Eq.~(\ref{eq:lin_eq}) for $\beta_1=0$ and $\beta_3=0$ can be specified in closed form \cite{Akhmediev:OC:1994,Tam:PRA:2020}, we here solve Eq.~(\ref{eq:lin_eq}) numerically in the range $\kappa\in (0, 2.5)~\mathrm{km^{-1}}$.
%
Considering the paramters listed in Tab.~\ref{tab:table01} under label A, above a threshold value of $\kappa=3 \beta_2^2/(2|\beta_4|)=1.5~\mathrm{km^{-1}}$ \cite{Akhmediev:OC:1994,Tam:PRA:2020}, both roots form complex conjugate pairs $\lambda_1(\kappa)=\lambda_2^*(\kappa)$ [Fig.~\ref{fig:02}(a), see curves labeled A; cf.\ Fig.~1 of Ref.~\cite{Akhmediev:OC:1994}]. Below this threshold, they are purely real and differ in value. For decreasing wavenumber $\kappa\to 0$ we can expect the behavior of the localized solution to Eq.~(\ref{eq:HONSE}) to be determined by $\lambda_2$, i.e.\ the root with the smallest negative real part, providing the correct limiting behavior $\lambda(\kappa) = -\sqrt{2\kappa/|\beta_2|}$ of the fundamental NSE soliton [Fig.~\ref{fig:02}(a), see short dashed line]. While the real part of $\lambda_2$ specifies the exponential decay of the tails, its imaginary part determines the frequency offset at which the solution exists. 
Let us emphasize that for the parameters listed in Tab.~\ref{tab:table01} under A, our numerical results reproduce those of Ref.~\cite{Karlsson:OC:1994}.
%
%
Now, for the parameters considered here, listed in Tab.~\ref{tab:table01} under B, the exact SW solution found in Ref.~\cite{Kruglov:PRA:2018} has wavenumber $\kappa\approx 0.618~\mathrm{km^{-1}}$, and velocity $v\approx-2.182~\mathrm{km/ps}$ in the retarded frame of reference. Solutions corresponding to Eq.~(\ref{eq:lin_eq}) with $\beta_1=1/v$ indicate that, as suggested above, stationary nontrivial solutions exist only beyond $\kappa_0 = 0.112~\mathrm{km^{-1}}$ [Fig.~\ref{fig:02}(a), see curves labeled B].
Also in this case there exists a threshold value at $\kappa\approx 1.260~\mathrm{km^{-1}}$, above which the real parts of the roots $\lambda_{1,2}$ are degenerate. In this case, however, they do not form a complex conjugate pair. Below this threshold the roots are purely real and satisfy $\mathrm{Re}[\lambda_1]<\mathrm{Re}[\lambda_2]$.
Specifically, at $\kappa^\prime\approx 0.847~\mathrm{km^{-1}}$, i.e.\ the wavenumber of the KH SW in the reference frame in which it is stationary, we find $\lambda_2=( -1.449 - i\,0.500 )~\mathrm{ps^{-1}}$. 
Let us point out the excellent agreement of $\mathrm{Re}[\lambda_2]$ with the asymptotic decay of the true SW, given by $\mathrm{sech}^2(w \tau)\propto e^{-2w\tau}$, for $\tau\to \infty$, with $-2w \approx -1.450~\mathrm{ps^{-1}}$ [see filled circle in Fig.~\ref{fig:02}(a)]. 
Further, the value of $\mathrm{Im}[\lambda_2]$ agrees with the corresponding value in the exact SW solution Eq.~(\ref{eq:KH_SW}), given by $-\delta=-0.5~\mathrm{ps^{-1}}$ [see open circle in Fig.~\ref{fig:02}(a)]. 
(For completeness, the features of the $\mathrm{sech}^2$-solution of Ref.~\cite{Karlsson:OC:1994} for the parameters listed under label A in Tab.~\ref{tab:table01}, given by $-2w=-12/5~\mathrm{ps^{-1}}$ and $-\delta=0$ at $\kappa=24/25~\mathrm{km^{-1}}$, are shown by the open and filled squares in Fig.~\ref{fig:02}(a).)

\begin{table}[b!]
\caption{\label{tab:table01}
Parameters corresponding to cases labeled A and B in Fig.~\ref{fig:02}.
Columns from left to right: Labels, and parameters specifying the inverse GV, GVD, 3OD, and 4OD. 
}
\begin{ruledtabular}
\begin{tabular}{lcccc}
   & $\beta_1$ & $\beta_2$ & $\beta_3$ & $\beta_4$\\
 \textrm{Label}   & $\mathrm{(ps/km)}$ & $\mathrm{(ps^2/km)}$ & $\mathrm{(ps^3/km)}$ & $\mathrm{(ps^4/km)}$ \\
\colrule
\textrm{A} & 0 & -1 & 0 & -1\\
\textrm{B} & $1/v\approx -0.458$ & -1 & 1/2 & -1\\
\end{tabular}
\end{ruledtabular}
\end{table}

\paragraph*{Analysis of localized states obtained by the SRM.}
So as to assess in which way the shape of full SW solutions to Eq.~(\ref{eq:HONSE}) depend on their wavenumber, we performed a parameter study keeping the inverse group velocity of the retarded frame of reference fixed at $\beta_1=1/v\approx -0.458~\mathrm{km/ps}$.
We then employed the SRM to retrieve localized states for a sequence of wavenumbers $\kappa>\kappa_0$. For the resulting family of SWs we  performed a fit to the three-parameter model $A^{\rm{fit}}(\tau)=A_0 \,\mathrm{sech}^\nu(\tau/\tau_0)$ in order to retrieve the pulse peak amplitude ($A_0$), pulse duration ($\tau_0$), and shape parameter ($\nu$) via parameter fitting to $A = |\psi|$.
Exemplary results of the fitting procedure for three selected wavenumbers are shown in Fig.~\ref{fig:01}(c) alongside the numerically exact SRM results. 
In summary, we found the pulse amplitude and duration to satisfy the empirical scaling relation
\begin{align}
A_0(\tau_0) = 0.12 - \frac{0.43}{\tau_0} + \frac{5.47}{\tau_0^2} -\frac{8.63}{\tau_0^3} + \frac{6.45}{\tau_0^4}, \label{eq:A0_tau0_fit} 
\end{align}
while pulse amplitude and wavenumber satisfy
\begin{align}
A_0(\kappa)=1.36 \,(\kappa-0.113)^{0.48}. \label{eq:A0_fit}
\end{align}
For clarity, the physical units of the parameters in Eqs.~(\ref{eq:A0_tau0_fit}-\ref{eq:A0_fit}) are suppressed.
%
%
While coming very close to the scaling behavior $A_0(\kappa)=\sqrt{2\kappa/\gamma}$, expected for fundamental solitons of the standard NSE, Eq.~(\ref{eq:A0_fit}) indicates smaller peak amplitudes for the considered localized states.
In this regard, we found that the linear dependence of pulse amplitude on inverse pulse duration, characteristic for the NSE, can be recovered in the limit of small wavenumbers where the localized state has a very narrow spectrum. This can be seen in Fig.~\ref{fig:02}(b), where the SRM results (solid line labeled B) are compared to $A_0(\tau_0) = \sqrt{|\beta_2^\prime|/(\gamma \tau_0^2)}$ (dashed line labeled NSE) with the value of GVD taken at the frequency offset of the localized state, i.e.\
$\beta_2^\prime \equiv \partial_\Omega^2\beta^\prime(\Omega)|_{\Omega=\delta}= -0.875\,\mathrm{ps^2/km}$.
This naive limit finds further support by noting that the pulse shape parameter approaches unity as $\kappa\to \kappa_0$ [Fig.~\ref{fig:02}(c)], indicative of the $\mathrm{sech}$-shape of the fundamental NSE soliton.
%
%
As evident from the energy-wavenumber diagram in Fig.~\ref{fig:02}(d), the considered localized states exhibit larger energy, and, due to Eq.~(\ref{eq:A0_fit}), lower peak intensity than fundamental solitons of the standard NSE at any given values of $\kappa$. 
As $\kappa\to \kappa_0$, the pulse energy also satisfies the naive NSE limit $E(\kappa)=\sqrt{8|\beta_2^\prime| (\kappa-\kappa_0)}/\gamma$. 
%
%
%
Let us note that we found the above three-paramter model to approximate the localized states obtained by the SRM very well for $\kappa_0 < \kappa < 0.9~\mathrm{km^{-1}}$. For wavenumbers $\kappa> 0.9~\mathrm{km^{-1}}$, the fit-model does not provide the correct asymptotics, but describes the central part of the pulses reasonably well [Fig.~\ref{fig:01}(c)].


\begin{figure}[t!]
\includegraphics[width=\linewidth]{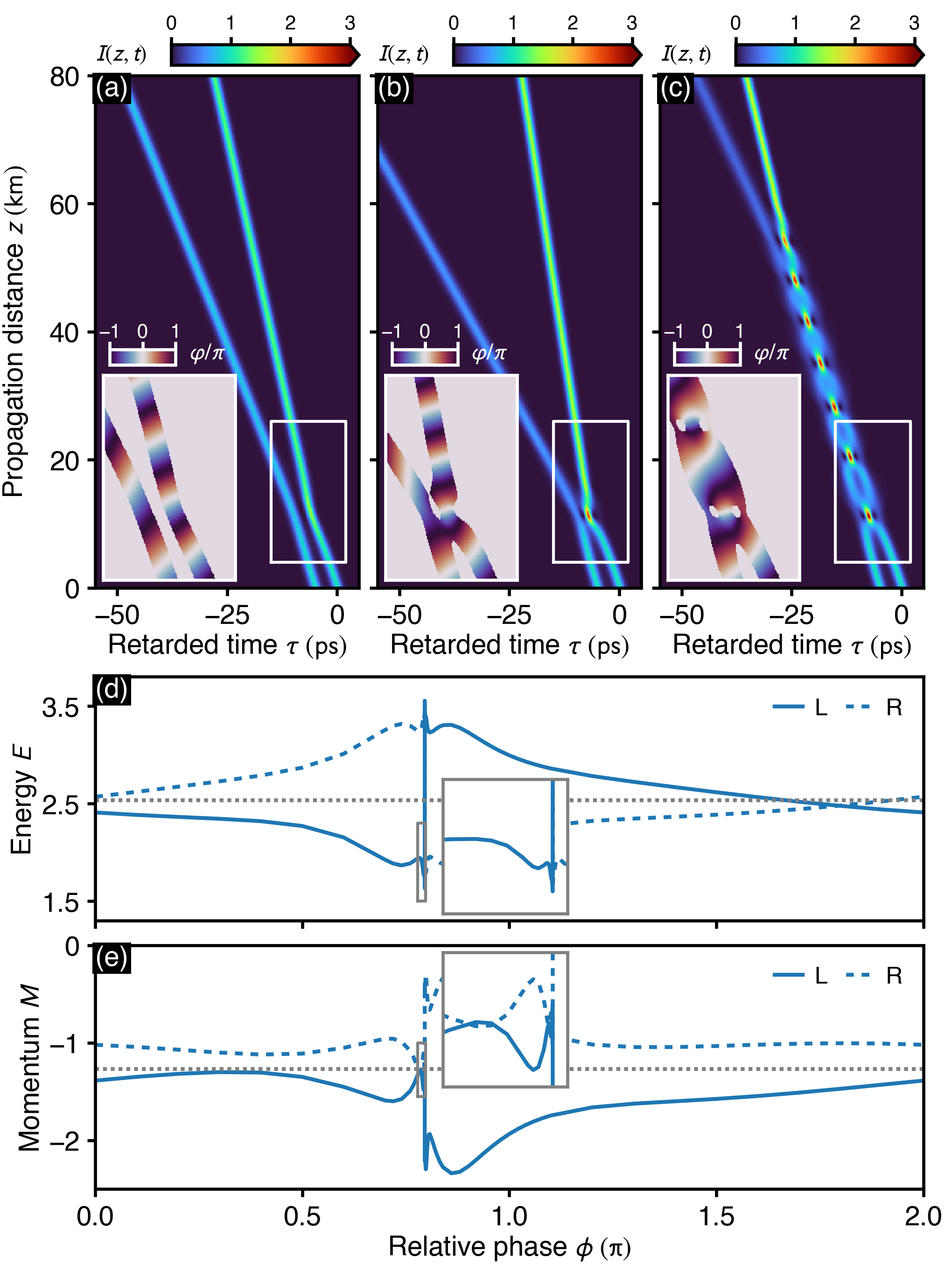}
\caption{Interaction dynamics of two KH solitary waves, depending on their relative phase $\phi$ at $z=0$.
(a) Time-domain propagation dynamics showing $I(z,t)=|\psi(z,t)|^2/\max(|\psi(0,t)|^2)$ for $\phi/\pi = 0.533$. The zoom-in shows a close up view of the pulse phase $\varphi(z,\tau)$ in the vicinity of their collision.
(b,c) Same as (a) for $\phi/\pi=0.733$, and $\phi/\pi=0.79577443$, respectively.
(d) Energy, and, (e) momentum of the left (L) and right (R) localized pulses at $z=80~\mathrm{km}$. 
In (d-e) short-dashed lines indicate energy and momentum of the KH SW and the zoom-in shows the range $\phi/\pi \in (0.779,0.798)$.
}
\label{fig:03a}
\end{figure}

\subsection{Collision dynamics involving the KH SW \label{sec:res02}}

As for other systems, which are described by perturbed variants of the NSE \cite{Anderson:PS:1986,Karpman:PD:1981,Yang:PRL:2000,Dmitriev:Chaos:2002,Dmitriev:PRE:2002,Frauenkron:PRE:1996,Melchert:SR:2021}, we expect the collision of the specific KH SW with an independent SW of Eq.~(\ref{eq:HONSE}) to exhibit diverse outcomes that depend on their mismatch in wavenumber, velocity, and initial phase.
Subsequently, we employ the SRM to obtain an independent SW solution $\psi^\prime(\tau)$ of Eq.~(\ref{eq:HONSE}) for a given wavenumber $\kappa^\prime$ and given inverse velocity $v^{\prime-1}$, and perform pulse propagation simulations for the initial condition 
\begin{align}
    \psi_0(\tau) = \psi^\prime(\tau-\tau_0^\prime) + \psi_{\rm{KH}}(0,\tau), \label{eq:ic_collision}
\end{align}
with $\psi_{\rm{KH}}$ defined by Eq.~(\ref{eq:KH_SW}) for $\eta=0$. Therein, our aim is to assess the dependence of the collision process on the initial phase $\phi$ of the KH SW, entering through Eq.~(\ref{eq:KH_SW}).

\paragraph*{Collision of two KH SWs.}

Considering $v^{\prime}=v$, and $\kappa^{\prime}=\kappa-\delta/v$, with $v$, $\delta$, and $\kappa$ determined by Eqs.~(\ref{eq:KH_SOl_v}-\ref{eq:KH_SOl_kappa}), yields an initial condition with two KH SWs at initial delay $\tau_0^\prime$. 
The smaller the initial delay, the larger the initial overlap between both pulses. Subsequently we consider $\tau_0^\prime=5~\mathrm{ps}$ and perform a parameter sweep by varying the initial phase $\phi$ of the right KH SWs. The results of our numerical simulations are summarized in Fig.~\ref{fig:03a}.
In Fig.~\ref{fig:03a}(a), the propagation dynamics at $\phi/\pi= 0.533$ is shown, demonstrating an inelastic
collision process in which energy is transferred to the right pulse [Fig.~\ref{fig:03a}(d)]. During an initial propagation stage, i.e.\ for $z<10~\mathrm{km}$, there exists an attractive interaction between both pulses. However, owing to the transfer of energy which is enabled by their initial phase difference, a wavenumber mismatch between both pulses quickly builds up, causing them to repel each other for $z\gtrapprox 10~\mathrm{km}$.
The pulse phase $\varphi(z,\tau) = \tan^{-1}\left( \mathrm{Im}[\psi] / \mathrm{Re}[\psi] \right)$, shown in the  
zoom-in in Fig.~\ref{fig:03a}(a), indicates that at the point of closest proximity, both pulses are indeed approximately out of phase.
When increasing the initial phase difference [Figs.~\ref{fig:03a}(b,c)], we find a narrow parameter range centered about $\phi/\pi \approx 0.796$, in which collisions are of ``in-phase'' type [see zoom-in in Fig.~\ref{fig:03a}(c)], where both pulses form a two-pulse bound state that decays after a finite number of collision events \cite{Dmitriev:Chaos:2002,Dmitriev:PRE:2002}. Therein, the number of collision events depends very sensibly on the initial phase. For instance, at $\phi/\pi = 0.79577443$, after a first collision at $z\approx 12~\mathrm{km}$, the resulting two-pulse bound state persists for 5 more collision events until it finally decays [Fig.~\ref{fig:03a}(c)]. 
Let us note that the oscillation period of the bound state, i.e.\ the $z$-separation between two successive collision events, decreases with increasing propagation distance. This is in contrast to the two-soliton bound states studied in terms of a quasi-particle approach in the standard NSE \cite{Anderson:PS:1986,Karpman:PD:1981}, or the weakly perturbed NSE \cite{Dmitriev:Chaos:2002}, which where found to oscillate with a fixed period.
Further, for $\phi/\pi \gtrapprox 0.796$, the collision dynamics is similar to Figs.~\ref{fig:03a}(a-c), only with a transfer of energy to the left pulse.
While ``in-phase'' type collisions are obtained for initial conditions at $\phi/\pi \approx 0.796$, ``out-of-phase'' type collisions are obtained at $\phi/\pi \approx 1.80$. 
As evident from the resulting energies [Fig.~\ref{fig:03a}(d)] and momenta [Fig.~\ref{fig:03a}(e)] at $z=80~\mathrm{km}$, none of the resulting pulses exhibit energy and momentum of a KH SW for any of the phase-values considered in our numerical experiments.
Similar to previous studies of soliton collisions in the weakly perturbed NSE \cite{Dmitriev:PRE:2002,Dmitriev:Chaos:2002,Frauenkron:PRE:1996}, we find that energy loss to free radiation is small. Only in the range of in-phase collisions, i.e.\ when two-soliton bound states are formed, an energy fraction of up to $\approx 6\%$ is converted to free radiation. 
Such a correlation between relative phase and radiation losses in SW collisions has earlier been observed for the saturable NSE \cite{Jakubowski:PRE:1997}.  
We further observe that summary measures, such as energy or momentum, exhibit self-similar features if viewed on different scales. Compare, e.g., the energy in range $\phi/\pi \in (0.779, 0.798)$, shown in the inset in Fig.~\ref{fig:03a}(d), to the energy in range $\phi/\pi\in (0.3, 0.8)$. A similar behavior has been observed for the postcollision properties of solitons in symmetric collisions in Refs.~\cite{Yang:PRL:2000,Dmitriev:PRE:2002}.  

\begin{figure}[t!]
\includegraphics[width=\linewidth]{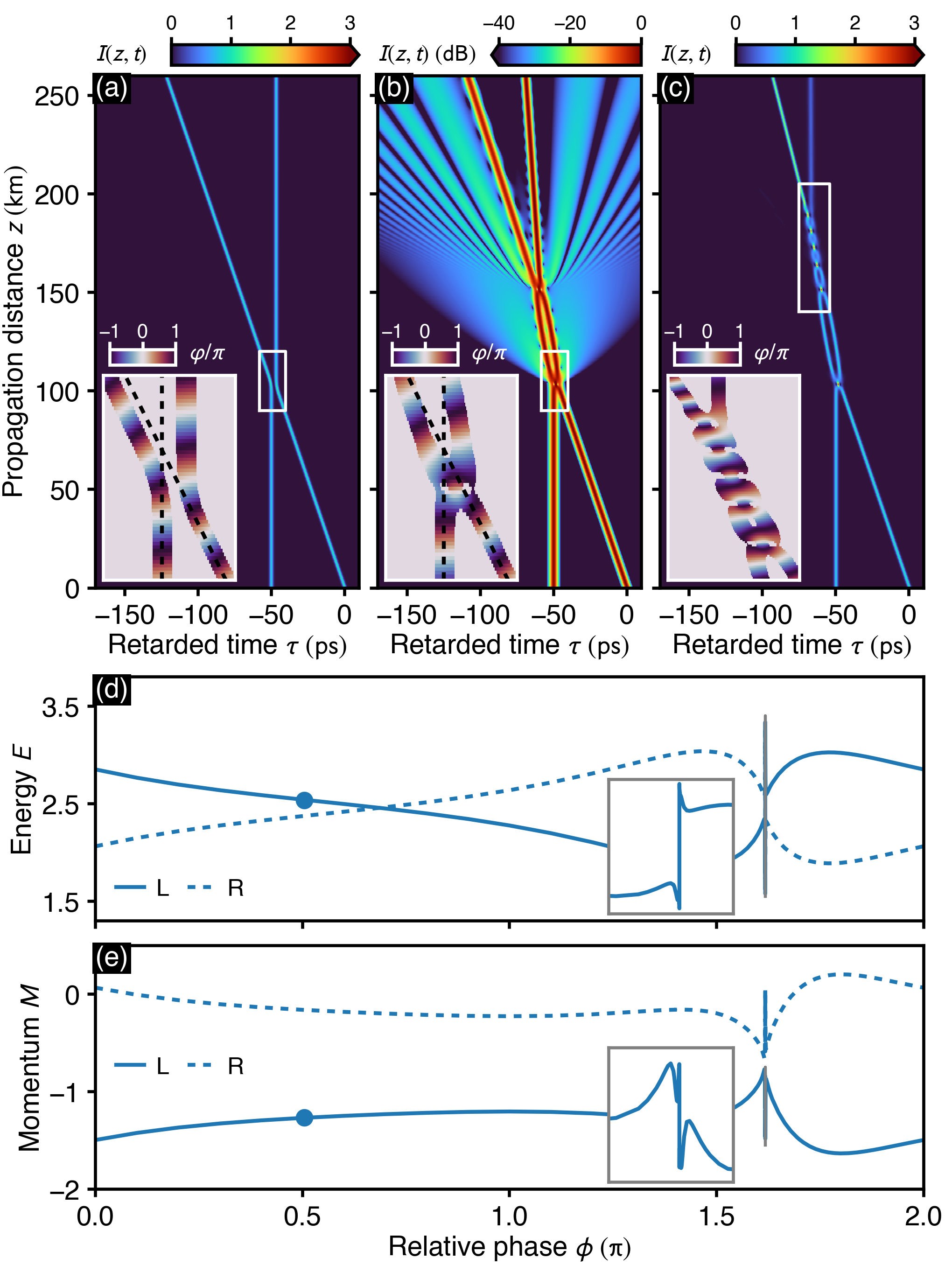}
\caption{Collision of a KH solitary wave and a stationary SW with matching wavenumbers, depending on their relative phase $\phi$ at $z=0$.
(a) Time-domain propagation dynamics showing $I(z,t)=|\psi(z,t)|^2/\max(|\psi(0,t)|^2)$ for $\phi/\pi = 0.505$. The zoom-in shows a close up view of the pulse phase $\varphi(z,\tau)$ in the vicinity of their collision. Dashed lines indicate reference trajectories with inverse velocity $0~\mathrm{ps/km}$ and $v^{-1}\approx -0.458~\mathrm{ps/km}$.
%
(b) Same as (a) for $\phi/\pi=0.168$, using a dB-scale.
(c) Same as (a) for $\phi/\pi=1.61804225$.
(d) Energy, and, (e) momentum of the left (L) and right (R) localized pulses at $z=250~\mathrm{km}$. 
In (d-e) the zoom-in shows the range $\phi/\pi \in (1.61790,1.81815)$.
}
\label{fig:03}
\end{figure}

\begin{figure}[t!]
\includegraphics[width=\linewidth]{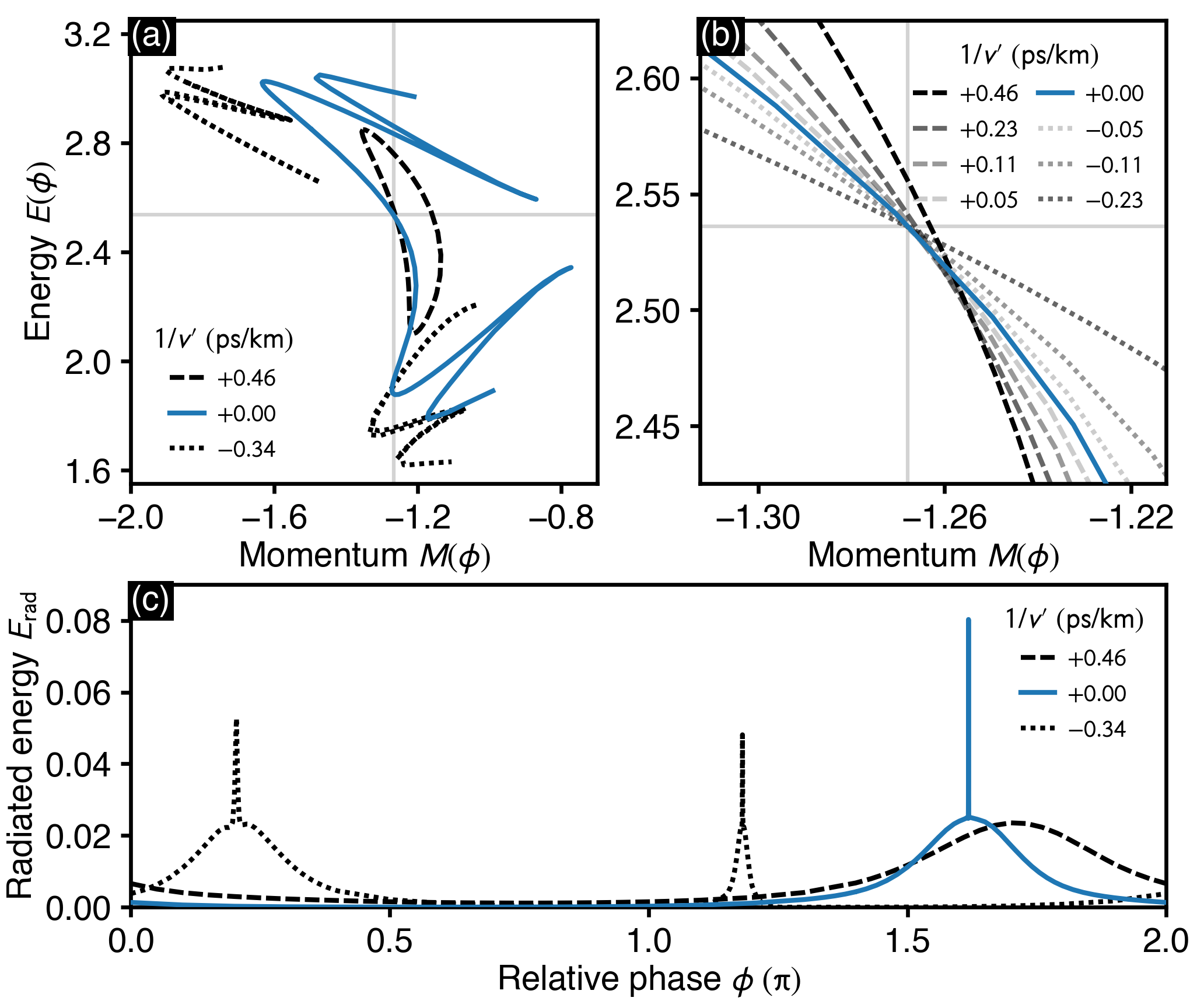}
\caption{Post-collision properties of the fast SW in the energy-momentum plane.
(a) Different scenarios observed for three selected inverse velocities $1/v^{\prime}$. 
(b) Zoom-in on a narrow energy-momentum range enclosing the values $E_{\rm{KH}}=2.536~\mathrm{W\,ps}$ and $M_{\rm{KH}}=-1.268~\mathrm{W}$, characterizing the KH SW.
(c) Fraction of total energy $E_{\mathrm{rad}}$ lost to radiation as function of initial phase $\phi$ for the collision scenarios in (a).
}
\label{fig:03c}
\end{figure}

\paragraph*{Collision at large relative velocities.}
In the sector of nonzero velocity mismatch, we expect complex interaction processes to occur for $\kappa^\prime \approx \kappa$. Below we report our results for $v^{\prime-1}=0~\mathrm{ps/km}$ and $\kappa^\prime = \kappa = 0.618~\mathrm{km^{-1}}$, for which the SRM yields a localized solution that can be parameterized as $\psi^\prime(\tau)= A_0 \,{\rm{sech}}^{\nu}(\tau/\tau_0)\,e^{-i b \tau}$ with $A_0 = 1.087~\mathrm{W^{1/2}}$, $\tau_0=1.388~\mathrm{ps}$, $\nu=1.728$, $b=0.07~\mathrm{ps^{-1}}$, and initial energy and momentum given by $E=2.378~\mathrm{W\,ps}$ and $M=-0.166~\mathrm{W}$.
Our results for the initial separation $\tau_0^\prime=50~\mathrm{ps}$ and initial relative phases in the range $\phi \in (0,2 \pi)$ in $\psi_{\rm{KH}}$ [see Eq.~(\ref{eq:KH_SW})] are summarized in Fig.~\ref{fig:03}.
We find that at $\phi/\pi \approx 0.505$ [Fig.~\ref{fig:03}(a)], the SWs engange in an out-of-phase type elastic collision after which the amplitude, shape, and velocity of both pulses are fully restored. 
As can be seen from the pulse phase $\varphi(z,\tau)$ [zoom-in in Fig.~\ref{fig:03}(a)], at the point of closest proximity, both pulses are indeed out ouf phase, and the effect of their mutual interaction is simply a shift of their respective loci relative to their free propagation. 
Let us point out that the post-collision energy and momentum of the left pulse at $z=250~\mathrm{km}$ matches the energy and momentum of the KH SW, indicated by the circles at $\phi/\pi=0.505$ in Figs.~\ref{fig:03}(d,e), with a negligible fraction on the order of $10^{-5}$ of the total energy converted to free radiation.
Thus, at this value of $\phi$, the collision is elastic.
Further, centered around $\phi/\pi \approx 1.6$ we find a narrow window within which two-pulse bound states are formed. As above, their $z$-lifetime is sensitive to their initial phase, and the $z$-oscillation period decreases between successive collision events.
Specifically, Fig.~\ref{fig:03}(b) shows such a bound state for $\phi/\pi=0.168$, which, after the first collision at $z\approx 100~\mathrm{km}$, persists for one more collision event, and Fig.~\ref{fig:03}(c) shows the propagation dynamics at $\phi/\pi=0.161804225$, exhibiting a two-pulse bound state that decays after 6 collision events. 
%
Bound-state formation with quite similar features have been observed during inelastic in-phase collisions of dipolar solitons at intermediate dipole-dipole interaction strengths \cite{Edmonds:NJP:2017}.
Let us point out that the time-domain intensity in Fig.~\ref{fig:03}(b) is shown on a dB-scale to demonstrate the intensity level on which free radiation is produced ($<-20~\mathrm{dB}$). In this case, the fraction of energy transferred to free radiation amounts to $\approx 0.04$. 
In Fig.~\ref{fig:03}(c), a fraction of $\approx 0.08$ of the total energy is converted to free radiation [see Fig.~\ref{fig:03c}(c); $v^{\prime-1}=0$].
In particular, in the zoom-in in Fig.~\ref{fig:03}(b) it can be seen that the free radiation, trapped in between the two pulses beyond their first collision at $z\approx 100~\mathrm{km}$, is in-phase with the trailing edge of the left pulse as well as the leading edge of the right pulse. This imparts a net attraction between both pulses, causing them to collide again at $z\approx150~\mathrm{km}$. 
%
%
%
While the considered propagation dynamics unfolds entirely at anomalous dispersion, a quite similar effect, involving two solitons cross-trapping a weak dispersive wave in a domain of normal dispersion, has been described in theory \cite{Anderson:EL:1992}, and observed in experiment \cite{Driben:OE:2013,Wang:OL:2015}.
Here, after the two-pulse bound states decay, none of the resulting pulses shares the properties of the KH SW for any of the parameters considered in our numerical experiments.

We assess the post-collision properties of the ``fast'' SW, i.e.\ the left SW after the collision events, in the energy-momentum plane in Figs.~\ref{fig:03c}(a,b)].
At large relative velocities [dashed line in Fig.~\ref{fig:03c}(a); $v^{\prime-1}=|v|^{-1}\approx 0.46~\mathrm{ps/km}$], possible energies are met for two distinct momenta. However, at no point in $\phi$ does the fast SW exhibit the exact properties of the KH SW [Fig.~\ref{fig:03c}(b)]. 
In addition, SWs at such a large initial velocity mismatch are immediately able to
overcome their mutual binding at in-phase collisions, inhibiting any bound-state formation.
At intermediate relative velocities [solid line in Fig.~\ref{fig:03c}(a); $v^{\prime-1}=0$], and for
initial phase differences enabling two-soliton bound systems, an energy gap at small negative momenta opens up wherein a unique relation between energy and momenta is given. Further, in the range of intermediate relative velocities, a unique value of the initial phase exist at which the KH SW continues to exist after the collision [Fig.~\ref{fig:03c}(a)].
Finally, at small relative velocities [short dashed line in Fig.~\ref{fig:03c}(a); $v^{\prime-1}\approx 0.74\,v^{-1}$], a second parameter range enabling two-soliton bound systems appears, leading to a total energy gap wherein no post-collision SW is found. Since the energy gap includes the energy $E_{\rm{KH}}=2.536~\mathrm{W\,ps}$, in this case, the KH SW persists at no value of the initial phase.
Examples of the amount of radiative losses observed subsequent to the mutual interaction of both pulses for three different values of the group-velocity $v^{\prime}$ are shown in Fig.~\ref{fig:03c}(c).


\subsection{Decay of the KH SW under absorption \label{sec:res03}}

\begin{figure}[t!]
\includegraphics[width=\linewidth]{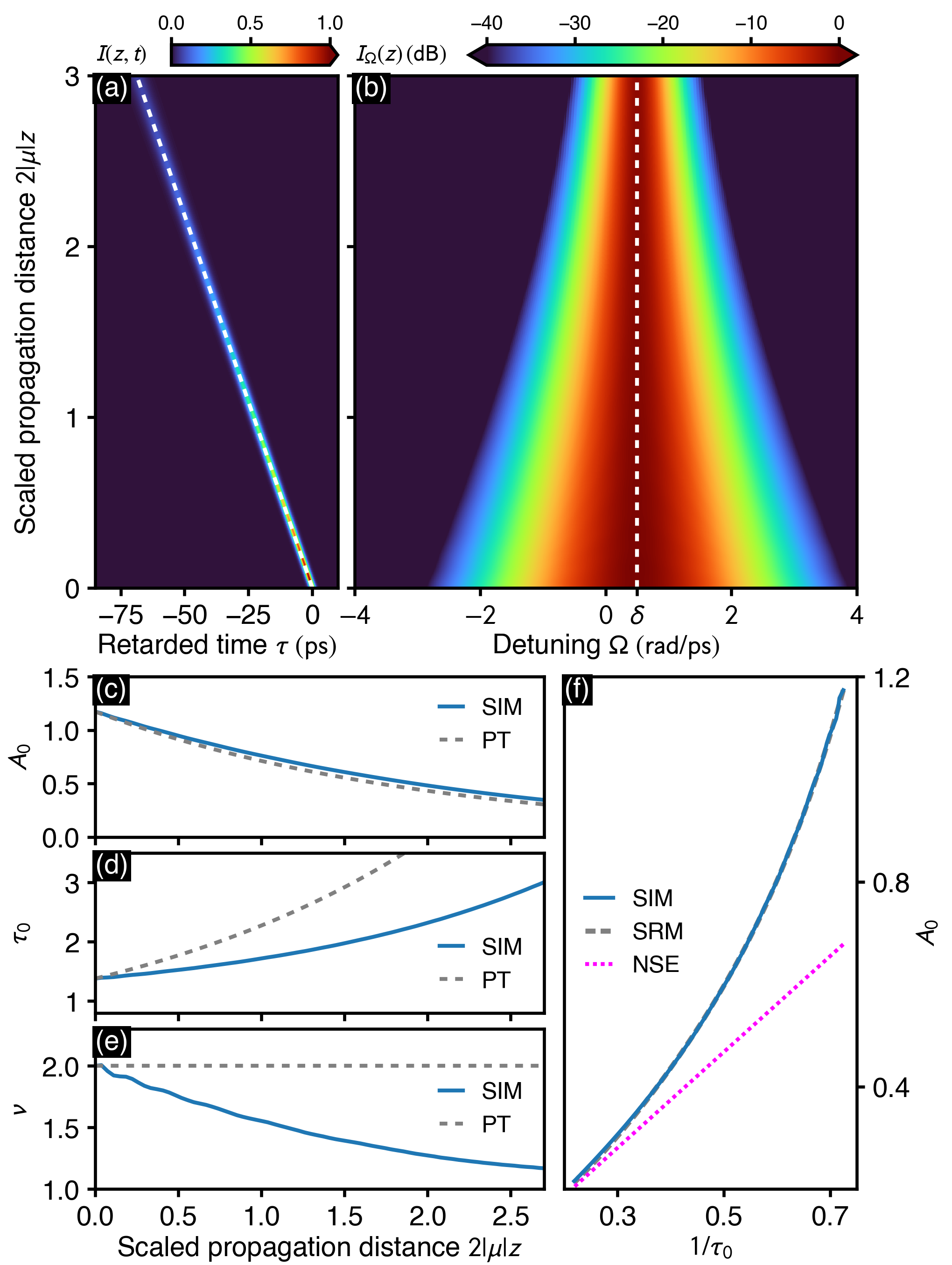}
\caption{Decay of the KH solitary wave in presence of absorption ($\mu=0.01\,\mathrm{km^{-1}}$).
(a) Time-domain propagation dynamics showing $I(z,t)=|\psi(z,t)|^2/\max(|\psi(0,t)|^2)$. White dashed line indicates a trajectory with reference 
velocity $v=-2.182\,\mathrm{km/ps}$. 
(b) Corresponding spectrum $I_\Omega(z)=|\psi_\Omega(z)|^2/\max(|\psi_\Omega(0)|^2)$. White dashed line indicates frequency offset $\delta=0.5\,\mathrm{rad/ps}$.
(c-e) Comparison of simulation results (SIM) and predictions based on perturbation theory (PT) \cite{Kruglov:PRA:2018}.
Variation of (c) amplitude, (d) duration, and, (e) pulse shape.
(f) Dependence of pulse amplitude on inverse temporal width.
Comparison of simulation results (SIM) to a fit-model based on SRM results (SRM), and limiting case of a standard NSE (NSE). Details are provided in the text.  
}
\label{fig:04}
\end{figure}

In an effort to go beyond the predictions of perturbation theory, obtained in Ref.~\cite{Kruglov:PRA:2018} to estimate the variation of the pulse properties under gain and loss for short propagation distances, we here perform pulse propagation simulations for the KH SW in terms of Eq.~(\ref{eq:HONSE}) for nonzero $\mu>0$.
%
%
Specifically, the considered example assumes an absorption coefficient $\mu = 0.01\,\mathrm{m^{-1}}$, yielding $|\mu \epsilon|/\alpha^2 =  0.00167 \ll 1$, in agreement with the conditions that enable comparison to the results reported in Ref.~\cite{Kruglov:PRA:2018}.
The propagation dynamics of the perturbed KH SW is shown in Figs.~\ref{fig:04}(a,b). As evident from Fig.~\ref{fig:04}(a), pulse intensity decreases and pulse duration increases for increasing propagation distance. Correspondingly, the extend of the pulse in the spectral domain decreases [Fig.~\ref{fig:04}(b)].
In order to analyze the shape of the perturbed KH SW upon propagation, we employ a four-parameter model of the form $A^{\rm{fit}}(\tau)=A_0\,\mathrm{sech}^\nu[(\tau-\tau_c)/\tau_0]$ and retrieve the pulse peak amplitude ($A_0$), peak position ($\tau_c$), duration ($\tau_0$), and shape parameter ($\nu$) through parameter fitting to $A(z,\tau) = |\psi(z,\tau)|$ at fixed $z$. The variation of the resulting parameters as function of the scaled propagation coordinate $2|\mu| z$ is shown in Figs.~\ref{fig:04}(c-e) (labeled SIM).
In the range $2|\mu|z\ll 1$, the predictions of the perturbation theory reported by Ref.~\cite{Kruglov:PRA:2018}, yielding amplitude $u^\prime = u\,e^{-\mu z}$ and duration ${w^\prime}^{-1} = w^{-1}\,e^{\mu z}$ [dashed lines in Fig.~\ref{fig:04}(c-e)], can be seen to agree very well with the simulation results.
Let us note that while the perturbative treatment yields an approximate solution with fixed $\nu=2$, our numerical results indicate that the shape parameter initially decreases $\propto 1 + e^{-\mu z}$ [Fig.~\ref{fig:04}(e)].
Moreover, based on the observation that the pulse acquires a rather narrow spectrum for increasing propagation distance, we can even formulate an approximate relation between pulse peak intensity $A_0$ and pulse duration $\tau_0$ [Fig.~\ref{fig:04}(f)], valid in the limit of large $z$.
Assuming that the pulse has a narrow spectrum, we might hypothesize that its dynamics are governed by the value of group-velocity dispersion at the frequency offset $\delta$, i.e.\ $\beta_2^\prime= -0.875\,\mathrm{ps^2/km}$, and that higher orders of dispersion have only a marginal impact on the overall dynamics. Resorting to a standard NSE then implies $A_0 = \sqrt{|\beta_2^\prime|/(\gamma \tau_0^2)}$. As evident from Fig.~\ref{fig:04}(f) this naive model (labeled NSE) is in excellent agreement with the observed data (labeled SIM) for small $1/\tau_0$, i.e.\ for large propagation distances $z$. 
%
This naive model finds further support from the pulse shape parameter, which approaches a value of $\nu=1$ at large $z$ [Fig.~\ref{fig:04}(e)], indicative of the hyperbolic-secant shape of the fundamental soliton of the NSE.
Moreover, the empirical relation Eq.~(\ref{eq:A0_tau0_fit}) for pulse amplitude and pulse duration, obtained from the SRM results in Sect.~\ref{sec:res01}, fits the parameters of the decaying SW very well [Fig.~\ref{fig:04}(f), curve labeled SRM].
Finally, by inverting the amplitude-wavelength relation Eq.~(\ref{eq:A0_fit}), we confirmed that the wavenumber of the decaying pulse shown in Fig.~\ref{fig:04}(a) extrapolates to $\kappa = 0.113~\mathrm{km^{-1}}\approx \kappa_0$ in the limit $z\to \infty$.

\section{Conclusions \label{sec:summary}}

In conclusion, we have performed a numerical study of SWs in the NSE perturbed by 3OD and negative 4OD. For a fixed set of parameters, satisfying the bounds of existence of the KH SW solution found by Ref.~\cite{Kruglov:PRA:2018}, we studied the asymptotic decay of solutions in terms of a linear auxiliary model and contrasted the considered setting with the case of vanishing 3OD, which admits the earlier Karlsson-H\"o\"ok SW solution \cite{Karlsson:OC:1994,Akhmediev:OC:1994}.
%
%
We further employed the SRM to retrieve a sequence of numerically exact SWs at fixed group-velocity for various wavenumbers $\kappa$, exceeding a wavenumber threshold $\kappa_0$ below which no nontrivial solutions exist. In our numerical experiments we reproduced the KH SW solution \cite{Kruglov:PRA:2018}, discussed the limiting case of $\kappa \to \kappa_0$ in which the soliton solution of the standard (unperturbed) NSE is approached, and obtained a quantitative relationship between amplitude and wavenumber of the SWs.

We then performed pulse propagation simulations to determine the interaction dynamics of two KH SW solutions at fixed initial delay as function of the initial phase difference of both pulses, finding that none of the postcollision pulses carries the identity of the KH SW. 
For collisions between a KH SW and an independent SW at large group-velocity mismatches but with both pulses having similar wavenumbers, a single value of the initial phase difference was found to exist for which one of the postcollision pulses has the identity of the KH SW.  
In either case, we found that in the parameter range where the collisions are of out-of-phase type, losses to radiation are negligible. In the range where in-phase type collisions occur, radiative losses can a make up for $\approx 10\%$ of the total energy. These findings are consistent with results obtained for SW collisions in different types of perturbed NSEs \cite{Jakubowski:PRE:1997,Frauenkron:PRE:1996}.
In this latter range of parameters, two-pulse bound states are formed that persist for a finite number of collision events before they decay into separate pulses. While such quasiparticles have been discussed earlier for the discrete NSE \cite{Dmitriev:Chaos:2002,Dmitriev:PRE:2002}, the progression from bound state formation to decay differs in the present case: the $z$-separation between between two successive collisions decreases.
Similar findings where reported for bright matter-wave solitons of dipolar Bose-Einstein condensates \cite{Edmonds:NJP:2017}.

Finally, we performed pulse propagation simulations including absorption to complement the predictions of the perturbation theory presented in Ref.~\cite{Kruglov:PRA:2018}. In the range of applicability of the perturbation theory, i.e.\ for short propagation distances, we verified the predicted trend of the pulse amplitude and pulse duration of a decaying KH SW. Further, in the limit of large propagation distances, a naive model suggested the decaying pulse to approach the fundamental soliton solution of the standard NSE. Upon propagation, the decaying pulse remains within the family of solutions characterized by the velocity given in Eq.~(\ref{eq:KH_SOl_v}).  

Our results shed further light on the properties and interaction dynamics of a family of localized wave solutions encompassing the exact Kruglov-Harvey solitary-wave solution. 

\section*{Acknowledgements}
This work was supported by the Deutsche Forschungsgemeinschaft (DFG) under Germany’s Excellence Strategy within the Cluster of Excellence PhoenixD (Photonics, Optics, and Engineering—Innovation Across Disciplines) [EXC 2122, Project No.\ 390833453].

\bibliography{references}

\begin{thebibliography}{69}%
\makeatletter
\providecommand \@ifxundefined [1]{%
 \@ifx{#1\undefined}
}%
\providecommand \@ifnum [1]{%
 \ifnum #1\expandafter \@firstoftwo
 \else \expandafter \@secondoftwo
 \fi
}%
\providecommand \@ifx [1]{%
 \ifx #1\expandafter \@firstoftwo
 \else \expandafter \@secondoftwo
 \fi
}%
\providecommand \natexlab [1]{#1}%
\providecommand \enquote  [1]{``#1''}%
\providecommand \bibnamefont  [1]{#1}%
\providecommand \bibfnamefont [1]{#1}%
\providecommand \citenamefont [1]{#1}%
\providecommand \href@noop [0]{\@secondoftwo}%
\providecommand \href [0]{\begingroup \@sanitize@url \@href}%
\providecommand \@href[1]{\@@startlink{#1}\@@href}%
\providecommand \@@href[1]{\endgroup#1\@@endlink}%
\providecommand \@sanitize@url [0]{\catcode `\\12\catcode `\$12\catcode
  `\&12\catcode `\#12\catcode `\^12\catcode `\_12\catcode `\%12\relax}%
\providecommand \@@startlink[1]{}%
\providecommand \@@endlink[0]{}%
\providecommand \url  [0]{\begingroup\@sanitize@url \@url }%
\providecommand \@url [1]{\endgroup\@href {#1}{\urlprefix }}%
\providecommand \urlprefix  [0]{URL }%
\providecommand \Eprint [0]{\href }%
\providecommand \doibase [0]{https://doi.org/}%
\providecommand \selectlanguage [0]{\@gobble}%
\providecommand \bibinfo  [0]{\@secondoftwo}%
\providecommand \bibfield  [0]{\@secondoftwo}%
\providecommand \translation [1]{[#1]}%
\providecommand \BibitemOpen [0]{}%
\providecommand \bibitemStop [0]{}%
\providecommand \bibitemNoStop [0]{.\EOS\space}%
\providecommand \EOS [0]{\spacefactor3000\relax}%
\providecommand \BibitemShut  [1]{\csname bibitem#1\endcsname}%
\let\auto@bib@innerbib\@empty
\bibitem [{\citenamefont {Zabusky}\ and\ \citenamefont
  {Kruskal}(1965)}]{Zabusky:PRL:1965}%
  \BibitemOpen
  \bibfield  {author} {\bibinfo {author} {\bibfnamefont {N.~J.}\ \bibnamefont
  {Zabusky}}\ and\ \bibinfo {author} {\bibfnamefont {M.~D.}\ \bibnamefont
  {Kruskal}},\ }\bibfield  {title} {\bibinfo {title} {Interaction of "solitons"
  in a collisionless plasma and the recurrence of initial states},\ }\href@noop
  {} {\bibfield  {journal} {\bibinfo  {journal} {Phys. Rev. Lett.}\ }\textbf
  {\bibinfo {volume} {15}},\ \bibinfo {pages} {240} (\bibinfo {year}
  {1965})}\BibitemShut {NoStop}%
\bibitem [{\citenamefont {Scott}(1979)}]{Scott:CC:1979}%
  \BibitemOpen
  \bibfield  {author} {\bibinfo {author} {\bibfnamefont {A.}~\bibnamefont
  {Scott}},\ }\bibfield  {title} {\bibinfo {title} {{Citation Classic --
  Soliton - new concept in applied science}},\ }\href@noop {} {\bibfield
  {journal} {\bibinfo  {journal} {Current Contents -- Engineering Technology \&
  Applied Science}\ }\textbf {\bibinfo {volume} {34}},\ \bibinfo {pages} {L12}
  (\bibinfo {year} {1979})}\BibitemShut {NoStop}%
\bibitem [{\citenamefont {Scott}\ \emph {et~al.}(1973)\citenamefont {Scott},
  \citenamefont {Chu},\ and\ \citenamefont {McLaughlin}}]{Scott:IEEE:1973}%
  \BibitemOpen
  \bibfield  {author} {\bibinfo {author} {\bibfnamefont {A.~C.}\ \bibnamefont
  {Scott}}, \bibinfo {author} {\bibfnamefont {F.~Y.~F.}\ \bibnamefont {Chu}},\
  and\ \bibinfo {author} {\bibfnamefont {D.~W.}\ \bibnamefont {McLaughlin}},\
  }\bibfield  {title} {\bibinfo {title} {{The soliton - A new concept in
  applied science}},\ }\href@noop {} {\bibfield  {journal} {\bibinfo  {journal}
  {Proc. IEEE}\ }\textbf {\bibinfo {volume} {61}},\ \bibinfo {pages} {144383}
  (\bibinfo {year} {1973})}\BibitemShut {NoStop}%
\bibitem [{\citenamefont {Zakharov}\ and\ \citenamefont
  {Shabat}(1972)}]{Zakharov:JETP:1972}%
  \BibitemOpen
  \bibfield  {author} {\bibinfo {author} {\bibfnamefont {V.~E.}\ \bibnamefont
  {Zakharov}}\ and\ \bibinfo {author} {\bibfnamefont {A.~B.}\ \bibnamefont
  {Shabat}},\ }\bibfield  {title} {\bibinfo {title} {{Exact theory of
  two-dimensional self-focusing and one-dimensional self-modulation of waves in
  nonlinear media}},\ }\href@noop {} {\bibfield  {journal} {\bibinfo  {journal}
  {Sov. Phys. JETP}\ }\textbf {\bibinfo {volume} {34}},\ \bibinfo {pages} {62}
  (\bibinfo {year} {1972})}\BibitemShut {NoStop}%
\bibitem [{\citenamefont {Aossey}\ \emph {et~al.}(1992)\citenamefont {Aossey},
  \citenamefont {Skinner}, \citenamefont {Cooney}, \citenamefont {Williams},
  \citenamefont {Gavin}, \citenamefont {Andersen},\ and\ \citenamefont
  {Lonngren}}]{Aossey:PRA:1992}%
  \BibitemOpen
  \bibfield  {author} {\bibinfo {author} {\bibfnamefont {D.~W.}\ \bibnamefont
  {Aossey}}, \bibinfo {author} {\bibfnamefont {S.~R.}\ \bibnamefont {Skinner}},
  \bibinfo {author} {\bibfnamefont {J.~L.}\ \bibnamefont {Cooney}}, \bibinfo
  {author} {\bibfnamefont {J.~E.}\ \bibnamefont {Williams}}, \bibinfo {author}
  {\bibfnamefont {M.~T.}\ \bibnamefont {Gavin}}, \bibinfo {author}
  {\bibfnamefont {D.~R.}\ \bibnamefont {Andersen}},\ and\ \bibinfo {author}
  {\bibfnamefont {K.~E.}\ \bibnamefont {Lonngren}},\ }\bibfield  {title}
  {\bibinfo {title} {{Properties of Soliton-Soliton collisions}},\ }\href@noop
  {} {\bibfield  {journal} {\bibinfo  {journal} {Phys. Rev. A}\ }\textbf
  {\bibinfo {volume} {45}},\ \bibinfo {pages} {2606} (\bibinfo {year}
  {1992})}\BibitemShut {NoStop}%
\bibitem [{\citenamefont {Kivshar}\ and\ \citenamefont
  {Agrawal}(2003)}]{Kivshar:BOOK:2003}%
  \BibitemOpen
  \bibfield  {author} {\bibinfo {author} {\bibfnamefont {Y.~S.}\ \bibnamefont
  {Kivshar}}\ and\ \bibinfo {author} {\bibfnamefont {G.~P.}\ \bibnamefont
  {Agrawal}},\ }\href@noop {} {\emph {\bibinfo {title} {Optical Solitons: From
  Fibers to Photonic Crystals}}}\ (\bibinfo  {publisher} {Academic Press},\
  \bibinfo {year} {2003})\BibitemShut {NoStop}%
\bibitem [{\citenamefont {Mitschke}(2016)}]{Mitschke:BOOK:2016}%
  \BibitemOpen
  \bibfield  {author} {\bibinfo {author} {\bibfnamefont {F.}~\bibnamefont
  {Mitschke}},\ }\href@noop {} {\emph {\bibinfo {title} {Fiber Optics: Physics
  and Technology}}}\ (\bibinfo  {publisher} {Springer},\ \bibinfo {year}
  {2016})\BibitemShut {NoStop}%
\bibitem [{\citenamefont {Russell}(1844)}]{Russell:BAR:1844}%
  \BibitemOpen
  \bibfield  {author} {\bibinfo {author} {\bibfnamefont {J.~S.}\ \bibnamefont
  {Russell}},\ }\bibfield  {title} {\bibinfo {title} {{Report on waves}},\
  }\href@noop {} {\bibfield  {journal} {\bibinfo  {journal} {Brit. Ass. Rep.}\
  }\textbf {\bibinfo {volume} {14}},\ \bibinfo {pages} {311} (\bibinfo {year}
  {1844})}\BibitemShut {NoStop}%
\bibitem [{\citenamefont {Jin}\ \emph {et~al.}(2024)\citenamefont {Jin},
  \citenamefont {Yang}, \citenamefont {Liao}, \citenamefont {Jing},\ and\
  \citenamefont {Yang}}]{Jin:PRB:2024}%
  \BibitemOpen
  \bibfield  {author} {\bibinfo {author} {\bibfnamefont {X.-W.}\ \bibnamefont
  {Jin}}, \bibinfo {author} {\bibfnamefont {Z.-Y.}\ \bibnamefont {Yang}},
  \bibinfo {author} {\bibfnamefont {Z.-M.}\ \bibnamefont {Liao}}, \bibinfo
  {author} {\bibfnamefont {G.}~\bibnamefont {Jing}},\ and\ \bibinfo {author}
  {\bibfnamefont {W.-L.}\ \bibnamefont {Yang}},\ }\bibfield  {title} {\bibinfo
  {title} {{Unveiling stable one-dimensional magnetic solitons in magnetic
  bilayers}},\ }\href@noop {} {\bibfield  {journal} {\bibinfo  {journal} {Phys.
  Rev. B}\ }\textbf {\bibinfo {volume} {109}},\ \bibinfo {pages} {014414}
  (\bibinfo {year} {2024})}\BibitemShut {NoStop}%
\bibitem [{\citenamefont {Gardner}\ \emph {et~al.}(1967)\citenamefont
  {Gardner}, \citenamefont {Greene}, \citenamefont {Kruskal},\ and\
  \citenamefont {Miura}}]{Gardner:PRL:1967}%
  \BibitemOpen
  \bibfield  {author} {\bibinfo {author} {\bibfnamefont {C.}~\bibnamefont
  {Gardner}}, \bibinfo {author} {\bibfnamefont {J.}~\bibnamefont {Greene}},
  \bibinfo {author} {\bibfnamefont {M.}~\bibnamefont {Kruskal}},\ and\ \bibinfo
  {author} {\bibfnamefont {R.}~\bibnamefont {Miura}},\ }\bibfield  {title}
  {\bibinfo {title} {{Method for solving the Korteweg-de Vries equation}},\
  }\href@noop {} {\bibfield  {journal} {\bibinfo  {journal} {Phys. Rev. Lett.}\
  }\textbf {\bibinfo {volume} {19}},\ \bibinfo {pages} {1095} (\bibinfo {year}
  {1967})}\BibitemShut {NoStop}%
\bibitem [{\citenamefont {Lax}(1968)}]{Lax:CPAM:1968}%
  \BibitemOpen
  \bibfield  {author} {\bibinfo {author} {\bibfnamefont {P.}~\bibnamefont
  {Lax}},\ }\bibfield  {title} {\bibinfo {title} {{Integrals of nonlinear
  equations of evolution and solitary waves}},\ }\href@noop {} {\bibfield
  {journal} {\bibinfo  {journal} {Comm. Pure Appl. Math.}\ }\textbf {\bibinfo
  {volume} {21}},\ \bibinfo {pages} {467} (\bibinfo {year} {1968})}\BibitemShut
  {NoStop}%
\bibitem [{\citenamefont {Ablowitz}\ \emph {et~al.}(1973)\citenamefont
  {Ablowitz}, \citenamefont {Kaup}, \citenamefont {Newell},\ and\ \citenamefont
  {Segur}}]{Ablowitz:PRL:1973}%
  \BibitemOpen
  \bibfield  {author} {\bibinfo {author} {\bibfnamefont {M.~J.}\ \bibnamefont
  {Ablowitz}}, \bibinfo {author} {\bibfnamefont {D.~J.}\ \bibnamefont {Kaup}},
  \bibinfo {author} {\bibfnamefont {A.~C.}\ \bibnamefont {Newell}},\ and\
  \bibinfo {author} {\bibfnamefont {H.}~\bibnamefont {Segur}},\ }\bibfield
  {title} {\bibinfo {title} {{Nonlinear-Evolution Equations of Physical
  Significance}},\ }\href@noop {} {\bibfield  {journal} {\bibinfo  {journal}
  {Phys. Rev. Lett.}\ }\textbf {\bibinfo {volume} {31}},\ \bibinfo {pages}
  {125} (\bibinfo {year} {1973})}\BibitemShut {NoStop}%
\bibitem [{\citenamefont {Ablowitz}\ \emph {et~al.}(1974)\citenamefont
  {Ablowitz}, \citenamefont {Kaup}, \citenamefont {Newell},\ and\ \citenamefont
  {Segur}}]{Ablowitz:SAM:1974}%
  \BibitemOpen
  \bibfield  {author} {\bibinfo {author} {\bibfnamefont {M.}~\bibnamefont
  {Ablowitz}}, \bibinfo {author} {\bibfnamefont {D.}~\bibnamefont {Kaup}},
  \bibinfo {author} {\bibfnamefont {A.}~\bibnamefont {Newell}},\ and\ \bibinfo
  {author} {\bibfnamefont {H.}~\bibnamefont {Segur}},\ }\bibfield  {title}
  {\bibinfo {title} {{The inverse scattering transform - Fourier analysis for
  nonlinear problems}},\ }\href@noop {} {\bibfield  {journal} {\bibinfo
  {journal} {Studies Appl. Math.}\ }\textbf {\bibinfo {volume} {53}},\ \bibinfo
  {pages} {249} (\bibinfo {year} {1974})}\BibitemShut {NoStop}%
\bibitem [{\citenamefont {Satsuma}\ and\ \citenamefont
  {Yajima}(1974)}]{Satsuma:PTP:1974}%
  \BibitemOpen
  \bibfield  {author} {\bibinfo {author} {\bibfnamefont {J.}~\bibnamefont
  {Satsuma}}\ and\ \bibinfo {author} {\bibfnamefont {N.}~\bibnamefont
  {Yajima}},\ }\bibfield  {title} {\bibinfo {title} {{Initial Value Problems of
  One-Dimensional Self-Modulation of Nonlinear Waves in Dispersive Media}},\
  }\href@noop {} {\bibfield  {journal} {\bibinfo  {journal} {Prog. Theor. Phys.
  (Suppl.)}\ }\textbf {\bibinfo {volume} {55}},\ \bibinfo {pages} {284}
  (\bibinfo {year} {1974})}\BibitemShut {NoStop}%
\bibitem [{\citenamefont {Kaup}(1975)}]{Kaup:JMP:1975}%
  \BibitemOpen
  \bibfield  {author} {\bibinfo {author} {\bibfnamefont {D.~J.}\ \bibnamefont
  {Kaup}},\ }\bibfield  {title} {\bibinfo {title} {{Exact quantization of the
  nonlinear Schrödinger equation}},\ }\href@noop {} {\bibfield  {journal}
  {\bibinfo  {journal} {J. Math. Phys.}\ }\textbf {\bibinfo {volume} {16}},\
  \bibinfo {pages} {2036} (\bibinfo {year} {1975})}\BibitemShut {NoStop}%
\bibitem [{\citenamefont {Kivshar}(1989)}]{Kivshar:RMP:1989}%
  \BibitemOpen
  \bibfield  {author} {\bibinfo {author} {\bibfnamefont {Y.~S.}\ \bibnamefont
  {Kivshar}},\ }\bibfield  {title} {\bibinfo {title} {{Dynamics of solitons in
  nearly integrable systems}},\ }\href@noop {} {\bibfield  {journal} {\bibinfo
  {journal} {Rev. Mod. Phys.}\ }\textbf {\bibinfo {volume} {61}},\ \bibinfo
  {pages} {763} (\bibinfo {year} {1989})}\BibitemShut {NoStop}%
\bibitem [{\citenamefont {Gordon}(1983)}]{Gordon:OL:1983}%
  \BibitemOpen
  \bibfield  {author} {\bibinfo {author} {\bibfnamefont {J.~P.}\ \bibnamefont
  {Gordon}},\ }\bibfield  {title} {\bibinfo {title} {Interaction forces among
  solitons in optical fibers},\ }\href@noop {} {\bibfield  {journal} {\bibinfo
  {journal} {Opt. Lett.}\ }\textbf {\bibinfo {volume} {8}},\ \bibinfo {pages}
  {596} (\bibinfo {year} {1983})}\BibitemShut {NoStop}%
\bibitem [{\citenamefont {Frauenkron}\ \emph {et~al.}(1996)\citenamefont
  {Frauenkron}, \citenamefont {Kivshar},\ and\ \citenamefont
  {Malomed}}]{Frauenkron:PRE:1996}%
  \BibitemOpen
  \bibfield  {author} {\bibinfo {author} {\bibfnamefont {H.}~\bibnamefont
  {Frauenkron}}, \bibinfo {author} {\bibfnamefont {Y.~S.}\ \bibnamefont
  {Kivshar}},\ and\ \bibinfo {author} {\bibfnamefont {B.~A.}\ \bibnamefont
  {Malomed}},\ }\bibfield  {title} {\bibinfo {title} {Multisoliton collisions
  in nearly integrable systems},\ }\href@noop {} {\bibfield  {journal}
  {\bibinfo  {journal} {Phys. Rev. E}\ }\textbf {\bibinfo {volume} {54}},\
  \bibinfo {pages} {R2244} (\bibinfo {year} {1996})}\BibitemShut {NoStop}%
\bibitem [{\citenamefont {Jakubowski}\ \emph {et~al.}(1997)\citenamefont
  {Jakubowski}, \citenamefont {Steiglitz},\ and\ \citenamefont
  {Squier}}]{Jakubowski:PRE:1997}%
  \BibitemOpen
  \bibfield  {author} {\bibinfo {author} {\bibfnamefont {M.~H.}\ \bibnamefont
  {Jakubowski}}, \bibinfo {author} {\bibfnamefont {K.}~\bibnamefont
  {Steiglitz}},\ and\ \bibinfo {author} {\bibfnamefont {R.}~\bibnamefont
  {Squier}},\ }\bibfield  {title} {\bibinfo {title} {{Information transfer
  between solitary waves in the saturable Schr\"odinger equation}},\
  }\href@noop {} {\bibfield  {journal} {\bibinfo  {journal} {Phys. Rev. E}\
  }\textbf {\bibinfo {volume} {56}},\ \bibinfo {pages} {7267} (\bibinfo {year}
  {1997})}\BibitemShut {NoStop}%
\bibitem [{\citenamefont {Anastassiou}\ \emph {et~al.}(1999)\citenamefont
  {Anastassiou}, \citenamefont {Segev}, \citenamefont {Steiglitz},
  \citenamefont {Giordmaine}, \citenamefont {Mitchell}, \citenamefont {Shih},
  \citenamefont {Lan},\ and\ \citenamefont {Martin}}]{Anastassiou:PRL:1999}%
  \BibitemOpen
  \bibfield  {author} {\bibinfo {author} {\bibfnamefont {C.}~\bibnamefont
  {Anastassiou}}, \bibinfo {author} {\bibfnamefont {M.}~\bibnamefont {Segev}},
  \bibinfo {author} {\bibfnamefont {K.}~\bibnamefont {Steiglitz}}, \bibinfo
  {author} {\bibfnamefont {J.~A.}\ \bibnamefont {Giordmaine}}, \bibinfo
  {author} {\bibfnamefont {M.}~\bibnamefont {Mitchell}}, \bibinfo {author}
  {\bibfnamefont {M.-F.}\ \bibnamefont {Shih}}, \bibinfo {author}
  {\bibfnamefont {S.}~\bibnamefont {Lan}},\ and\ \bibinfo {author}
  {\bibfnamefont {J.}~\bibnamefont {Martin}},\ }\bibfield  {title} {\bibinfo
  {title} {Energy-exchange interactions between colliding vector solitons},\
  }\href@noop {} {\bibfield  {journal} {\bibinfo  {journal} {Phys. Rev. Lett.}\
  }\textbf {\bibinfo {volume} {83}},\ \bibinfo {pages} {2332} (\bibinfo {year}
  {1999})}\BibitemShut {NoStop}%
\bibitem [{\citenamefont {Yang}\ and\ \citenamefont
  {Tan}(2000)}]{Yang:PRL:2000}%
  \BibitemOpen
  \bibfield  {author} {\bibinfo {author} {\bibfnamefont {J.}~\bibnamefont
  {Yang}}\ and\ \bibinfo {author} {\bibfnamefont {Y.}~\bibnamefont {Tan}},\
  }\bibfield  {title} {\bibinfo {title} {{Fractal Structure in the Collision of
  Vector Solitons}},\ }\href@noop {} {\bibfield  {journal} {\bibinfo  {journal}
  {Phys. Rev. Lett.}\ }\textbf {\bibinfo {volume} {85}},\ \bibinfo {pages}
  {3624} (\bibinfo {year} {2000})}\BibitemShut {NoStop}%
\bibitem [{\citenamefont {Dmitriev}\ and\ \citenamefont
  {Shigenari}(2002)}]{Dmitriev:Chaos:2002}%
  \BibitemOpen
  \bibfield  {author} {\bibinfo {author} {\bibfnamefont {S.~V.}\ \bibnamefont
  {Dmitriev}}\ and\ \bibinfo {author} {\bibfnamefont {T.}~\bibnamefont
  {Shigenari}},\ }\bibfield  {title} {\bibinfo {title} {{Short-lived
  two-soliton bound states in weakly perturbed nonlinear Schrödinger
  equation}},\ }\href@noop {} {\bibfield  {journal} {\bibinfo  {journal}
  {Chaos}\ }\textbf {\bibinfo {volume} {12}},\ \bibinfo {pages} {324} (\bibinfo
  {year} {2002})}\BibitemShut {NoStop}%
\bibitem [{\citenamefont {Dmitriev}\ \emph {et~al.}(2002)\citenamefont
  {Dmitriev}, \citenamefont {Semagin}, \citenamefont {Sukhorukov},\ and\
  \citenamefont {Shigenari}}]{Dmitriev:PRE:2002}%
  \BibitemOpen
  \bibfield  {author} {\bibinfo {author} {\bibfnamefont {S.~V.}\ \bibnamefont
  {Dmitriev}}, \bibinfo {author} {\bibfnamefont {D.~A.}\ \bibnamefont
  {Semagin}}, \bibinfo {author} {\bibfnamefont {A.~A.}\ \bibnamefont
  {Sukhorukov}},\ and\ \bibinfo {author} {\bibfnamefont {T.}~\bibnamefont
  {Shigenari}},\ }\bibfield  {title} {\bibinfo {title} {{Chaotic character of
  two-soliton collisions in the weakly perturbed nonlinear Schr\"odinger
  equation}},\ }\href@noop {} {\bibfield  {journal} {\bibinfo  {journal} {Phys.
  Rev. E}\ }\textbf {\bibinfo {volume} {66}},\ \bibinfo {pages} {046609}
  (\bibinfo {year} {2002})}\BibitemShut {NoStop}%
\bibitem [{\citenamefont {Feigenbaum}\ and\ \citenamefont
  {Orenstein}(2004)}]{Feigenbaum:OE:2004}%
  \BibitemOpen
  \bibfield  {author} {\bibinfo {author} {\bibfnamefont {E.}~\bibnamefont
  {Feigenbaum}}\ and\ \bibinfo {author} {\bibfnamefont {M.}~\bibnamefont
  {Orenstein}},\ }\bibfield  {title} {\bibinfo {title} {Colored solitons
  interactions: particle-like and beyond},\ }\href@noop {} {\bibfield
  {journal} {\bibinfo  {journal} {Opt. Express}\ }\textbf {\bibinfo {volume}
  {12}},\ \bibinfo {pages} {2193} (\bibinfo {year} {2004})}\BibitemShut
  {NoStop}%
\bibitem [{\citenamefont {Feigenbaum}\ and\ \citenamefont
  {Orenstein}(2005)}]{Feigenbaum:JOSAB:2005}%
  \BibitemOpen
  \bibfield  {author} {\bibinfo {author} {\bibfnamefont {E.}~\bibnamefont
  {Feigenbaum}}\ and\ \bibinfo {author} {\bibfnamefont {M.}~\bibnamefont
  {Orenstein}},\ }\bibfield  {title} {\bibinfo {title} {Coherent interactions
  of colored solitons via parametric processes: modified perturbation
  analysis},\ }\href@noop {} {\bibfield  {journal} {\bibinfo  {journal} {J.
  Opt. Soc. Am. B}\ }\textbf {\bibinfo {volume} {22}},\ \bibinfo {pages} {1414}
  (\bibinfo {year} {2005})}\BibitemShut {NoStop}%
\bibitem [{\citenamefont {Dingwall}\ \emph {et~al.}(2018)\citenamefont
  {Dingwall}, \citenamefont {Edmonds}, \citenamefont {Helm}, \citenamefont
  {Malomed},\ and\ \citenamefont {Öhberg}}]{Dingwall:NJP:2018}%
  \BibitemOpen
  \bibfield  {author} {\bibinfo {author} {\bibfnamefont {R.~J.}\ \bibnamefont
  {Dingwall}}, \bibinfo {author} {\bibfnamefont {M.~J.}\ \bibnamefont
  {Edmonds}}, \bibinfo {author} {\bibfnamefont {J.~L.}\ \bibnamefont {Helm}},
  \bibinfo {author} {\bibfnamefont {B.~A.}\ \bibnamefont {Malomed}},\ and\
  \bibinfo {author} {\bibfnamefont {P.}~\bibnamefont {Öhberg}},\ }\bibfield
  {title} {\bibinfo {title} {Non-integrable dynamics of matter-wave solitons in
  a density-dependent gauge theory},\ }\href@noop {} {\bibfield  {journal}
  {\bibinfo  {journal} {New Journal of Physics}\ }\textbf {\bibinfo {volume}
  {20}},\ \bibinfo {pages} {043004} (\bibinfo {year} {2018})}\BibitemShut
  {NoStop}%
\bibitem [{\citenamefont {Rao}\ \emph {et~al.}(2020)\citenamefont {Rao},
  \citenamefont {He}, \citenamefont {Kanna},\ and\ \citenamefont
  {Mihalache}}]{Rao:PRE:2020}%
  \BibitemOpen
  \bibfield  {author} {\bibinfo {author} {\bibfnamefont {J.}~\bibnamefont
  {Rao}}, \bibinfo {author} {\bibfnamefont {J.}~\bibnamefont {He}}, \bibinfo
  {author} {\bibfnamefont {T.}~\bibnamefont {Kanna}},\ and\ \bibinfo {author}
  {\bibfnamefont {D.}~\bibnamefont {Mihalache}},\ }\bibfield  {title} {\bibinfo
  {title} {{Nonlocal $M$-component nonlinear Schr\"odinger equations: Bright
  solitons, energy-sharing collisions, and positons}},\ }\href@noop {}
  {\bibfield  {journal} {\bibinfo  {journal} {Phys. Rev. E}\ }\textbf {\bibinfo
  {volume} {102}},\ \bibinfo {pages} {032201} (\bibinfo {year}
  {2020})}\BibitemShut {NoStop}%
\bibitem [{\citenamefont {Edmonds}\ \emph {et~al.}(2017)\citenamefont
  {Edmonds}, \citenamefont {Bland}, \citenamefont {Doran},\ and\ \citenamefont
  {Parker}}]{Edmonds:NJP:2017}%
  \BibitemOpen
  \bibfield  {author} {\bibinfo {author} {\bibfnamefont {M.~J.}\ \bibnamefont
  {Edmonds}}, \bibinfo {author} {\bibfnamefont {T.}~\bibnamefont {Bland}},
  \bibinfo {author} {\bibfnamefont {R.}~\bibnamefont {Doran}},\ and\ \bibinfo
  {author} {\bibfnamefont {N.~G.}\ \bibnamefont {Parker}},\ }\bibfield  {title}
  {\bibinfo {title} {{Engineering bright matter-wave solitons of dipolar
  condensates}},\ }\href@noop {} {\bibfield  {journal} {\bibinfo  {journal}
  {New Journal of Physics}\ }\textbf {\bibinfo {volume} {19}},\ \bibinfo
  {pages} {023019} (\bibinfo {year} {2017})}\BibitemShut {NoStop}%
\bibitem [{\citenamefont {Goodman}\ and\ \citenamefont
  {Haberman}(2007)}]{Goodman:PRL:2007}%
  \BibitemOpen
  \bibfield  {author} {\bibinfo {author} {\bibfnamefont {R.~H.}\ \bibnamefont
  {Goodman}}\ and\ \bibinfo {author} {\bibfnamefont {R.}~\bibnamefont
  {Haberman}},\ }\bibfield  {title} {\bibinfo {title} {{Chaotic Scattering and
  the $n$-Bounce Resonance in Solitary-Wave Interactions}},\ }\href@noop {}
  {\bibfield  {journal} {\bibinfo  {journal} {Phys. Rev. Lett.}\ }\textbf
  {\bibinfo {volume} {98}},\ \bibinfo {pages} {104103} (\bibinfo {year}
  {2007})}\BibitemShut {NoStop}%
\bibitem [{\citenamefont {Tsoy}\ and\ \citenamefont
  {de~Sterke}(2007)}]{Tsoy:PRA:2007}%
  \BibitemOpen
  \bibfield  {author} {\bibinfo {author} {\bibfnamefont {E.~N.}\ \bibnamefont
  {Tsoy}}\ and\ \bibinfo {author} {\bibfnamefont {C.~M.}\ \bibnamefont
  {de~Sterke}},\ }\bibfield  {title} {\bibinfo {title} {Theoretical analysis of
  the self-frequency shift near zero-dispersion points: Soliton spectral
  tunneling},\ }\href@noop {} {\bibfield  {journal} {\bibinfo  {journal} {Phys.
  Rev. A}\ }\textbf {\bibinfo {volume} {76}},\ \bibinfo {pages} {043804}
  (\bibinfo {year} {2007})}\BibitemShut {NoStop}%
\bibitem [{\citenamefont {Serkin}\ \emph {et~al.}(1993)\citenamefont {Serkin},
  \citenamefont {Vysloukh},\ and\ \citenamefont {Taylor}}]{Serkin:EL:1993}%
  \BibitemOpen
  \bibfield  {author} {\bibinfo {author} {\bibfnamefont {V.~N.}\ \bibnamefont
  {Serkin}}, \bibinfo {author} {\bibfnamefont {V.~A.}\ \bibnamefont
  {Vysloukh}},\ and\ \bibinfo {author} {\bibfnamefont {J.~R.}\ \bibnamefont
  {Taylor}},\ }\bibfield  {title} {\bibinfo {title} {{Soliton spectral
  tunnelling effect}},\ }\href@noop {} {\bibfield  {journal} {\bibinfo
  {journal} {Electron. Lett.}\ }\textbf {\bibinfo {volume} {29}},\ \bibinfo
  {pages} {12} (\bibinfo {year} {1993})}\BibitemShut {NoStop}%
\bibitem [{\citenamefont {Melchert}\ \emph {et~al.}(2019)\citenamefont
  {Melchert}, \citenamefont {Willms}, \citenamefont {Bose}, \citenamefont
  {Yulin}, \citenamefont {Roth}, \citenamefont {Mitschke}, \citenamefont
  {Morgner}, \citenamefont {Babushkin},\ and\ \citenamefont
  {Demircan}}]{Melchert:PRL:2019}%
  \BibitemOpen
  \bibfield  {author} {\bibinfo {author} {\bibfnamefont {O.}~\bibnamefont
  {Melchert}}, \bibinfo {author} {\bibfnamefont {S.}~\bibnamefont {Willms}},
  \bibinfo {author} {\bibfnamefont {S.}~\bibnamefont {Bose}}, \bibinfo {author}
  {\bibfnamefont {A.}~\bibnamefont {Yulin}}, \bibinfo {author} {\bibfnamefont
  {B.}~\bibnamefont {Roth}}, \bibinfo {author} {\bibfnamefont {F.}~\bibnamefont
  {Mitschke}}, \bibinfo {author} {\bibfnamefont {U.}~\bibnamefont {Morgner}},
  \bibinfo {author} {\bibfnamefont {I.}~\bibnamefont {Babushkin}},\ and\
  \bibinfo {author} {\bibfnamefont {A.}~\bibnamefont {Demircan}},\ }\bibfield
  {title} {\bibinfo {title} {Soliton molecules with two frequencies},\
  }\href@noop {} {\bibfield  {journal} {\bibinfo  {journal} {Phys. Rev. Lett.}\
  }\textbf {\bibinfo {volume} {123}},\ \bibinfo {pages} {243905} (\bibinfo
  {year} {2019})}\BibitemShut {NoStop}%
\bibitem [{\citenamefont {Melchert}\ and\ \citenamefont
  {Demircan}(2021)}]{Melchert:OL:2021}%
  \BibitemOpen
  \bibfield  {author} {\bibinfo {author} {\bibfnamefont {O.}~\bibnamefont
  {Melchert}}\ and\ \bibinfo {author} {\bibfnamefont {A.}~\bibnamefont
  {Demircan}},\ }\bibfield  {title} {\bibinfo {title} {Incoherent two-color
  pulse compounds},\ }\href@noop {} {\bibfield  {journal} {\bibinfo  {journal}
  {Opt. Lett.}\ }\textbf {\bibinfo {volume} {46}},\ \bibinfo {pages} {5603}
  (\bibinfo {year} {2021})}\BibitemShut {NoStop}%
\bibitem [{\citenamefont {Melchert}\ \emph {et~al.}(2023)\citenamefont
  {Melchert}, \citenamefont {Willms}, \citenamefont {Babushkin}, \citenamefont
  {Morgner},\ and\ \citenamefont {Demircan}}]{Melchert:OPTIK:2023}%
  \BibitemOpen
  \bibfield  {author} {\bibinfo {author} {\bibfnamefont {O.}~\bibnamefont
  {Melchert}}, \bibinfo {author} {\bibfnamefont {S.}~\bibnamefont {Willms}},
  \bibinfo {author} {\bibfnamefont {I.}~\bibnamefont {Babushkin}}, \bibinfo
  {author} {\bibfnamefont {U.}~\bibnamefont {Morgner}},\ and\ \bibinfo {author}
  {\bibfnamefont {A.}~\bibnamefont {Demircan}},\ }\bibfield  {title} {\bibinfo
  {title} {{(Invited) Two-color soliton meta-atoms and molecules}},\
  }\href@noop {} {\bibfield  {journal} {\bibinfo  {journal} {Optik}\ }\textbf
  {\bibinfo {volume} {280}},\ \bibinfo {pages} {170772} (\bibinfo {year}
  {2023})}\BibitemShut {NoStop}%
\bibitem [{\citenamefont {Karlsson}\ and\ \citenamefont
  {H\"o\"ok}(1994)}]{Karlsson:OC:1994}%
  \BibitemOpen
  \bibfield  {author} {\bibinfo {author} {\bibfnamefont {M.}~\bibnamefont
  {Karlsson}}\ and\ \bibinfo {author} {\bibfnamefont {A.}~\bibnamefont
  {H\"o\"ok}},\ }\bibfield  {title} {\bibinfo {title} {{Soliton-like pulses
  governed by fourth order dispersion in optical fibers}},\ }\href@noop {}
  {\bibfield  {journal} {\bibinfo  {journal} {Optics Communications}\ }\textbf
  {\bibinfo {volume} {104}},\ \bibinfo {pages} {303} (\bibinfo {year}
  {1994})}\BibitemShut {NoStop}%
\bibitem [{\citenamefont {Kruglov}\ and\ \citenamefont
  {Harvey}(2018)}]{Kruglov:PRA:2018}%
  \BibitemOpen
  \bibfield  {author} {\bibinfo {author} {\bibfnamefont {V.~I.}\ \bibnamefont
  {Kruglov}}\ and\ \bibinfo {author} {\bibfnamefont {J.~D.}\ \bibnamefont
  {Harvey}},\ }\bibfield  {title} {\bibinfo {title} {Solitary waves in optical
  fibers governed by higher-order dispersion},\ }\href@noop {} {\bibfield
  {journal} {\bibinfo  {journal} {Phys. Rev. A}\ }\textbf {\bibinfo {volume}
  {98}},\ \bibinfo {pages} {063811} (\bibinfo {year} {2018})}\BibitemShut
  {NoStop}%
\bibitem [{\citenamefont {Tam}\ \emph {et~al.}(2020)\citenamefont {Tam},
  \citenamefont {Alexander}, \citenamefont {Blanco-Redondo},\ and\
  \citenamefont {de~Sterke}}]{Tam:PRA:2020}%
  \BibitemOpen
  \bibfield  {author} {\bibinfo {author} {\bibfnamefont {K.~K.~K.}\
  \bibnamefont {Tam}}, \bibinfo {author} {\bibfnamefont {T.~J.}\ \bibnamefont
  {Alexander}}, \bibinfo {author} {\bibfnamefont {A.}~\bibnamefont
  {Blanco-Redondo}},\ and\ \bibinfo {author} {\bibfnamefont {C.~M.}\
  \bibnamefont {de~Sterke}},\ }\bibfield  {title} {\bibinfo {title}
  {{Generalized dispersion Kerr solitons}},\ }\href@noop {} {\bibfield
  {journal} {\bibinfo  {journal} {Phys. Rev. A}\ }\textbf {\bibinfo {volume}
  {101}},\ \bibinfo {pages} {043822} (\bibinfo {year} {2020})}\BibitemShut
  {NoStop}%
\bibitem [{\citenamefont {Lourdesamy}\ \emph {et~al.}(2021)\citenamefont
  {Lourdesamy}, \citenamefont {Runge}, \citenamefont {Alexander}, \citenamefont
  {Hudson}, \citenamefont {Blanco-Redondo},\ and\ \citenamefont
  {de~Sterke}}]{Lourdesamy:NP:2021}%
  \BibitemOpen
  \bibfield  {author} {\bibinfo {author} {\bibfnamefont {J.~P.}\ \bibnamefont
  {Lourdesamy}}, \bibinfo {author} {\bibfnamefont {A.~F.~J.}\ \bibnamefont
  {Runge}}, \bibinfo {author} {\bibfnamefont {T.~J.}\ \bibnamefont
  {Alexander}}, \bibinfo {author} {\bibfnamefont {D.~D.}\ \bibnamefont
  {Hudson}}, \bibinfo {author} {\bibfnamefont {A.}~\bibnamefont
  {Blanco-Redondo}},\ and\ \bibinfo {author} {\bibfnamefont {C.~M.}\
  \bibnamefont {de~Sterke}},\ }\bibfield  {title} {\bibinfo {title} {Spectrally
  periodic pulses for enhancement of optical nonlinear effects},\ }\href@noop
  {} {\bibfield  {journal} {\bibinfo  {journal} {Nat. Phys.}\ }\textbf
  {\bibinfo {volume} {18}},\ \bibinfo {pages} {59} (\bibinfo {year}
  {2021})}\BibitemShut {NoStop}%
\bibitem [{\citenamefont {Wai}\ \emph {et~al.}(1990)\citenamefont {Wai},
  \citenamefont {Chen},\ and\ \citenamefont {Lee}}]{Wai:PRA:1990}%
  \BibitemOpen
  \bibfield  {author} {\bibinfo {author} {\bibfnamefont {P.~K.~A.}\
  \bibnamefont {Wai}}, \bibinfo {author} {\bibfnamefont {H.~H.}\ \bibnamefont
  {Chen}},\ and\ \bibinfo {author} {\bibfnamefont {Y.~C.}\ \bibnamefont
  {Lee}},\ }\bibfield  {title} {\bibinfo {title} {Radiations by ``solitons'' at
  the zero group-dispersion wavelength of single-mode optical fibers},\
  }\href@noop {} {\bibfield  {journal} {\bibinfo  {journal} {Phys. Rev. A}\
  }\textbf {\bibinfo {volume} {41}},\ \bibinfo {pages} {426} (\bibinfo {year}
  {1990})}\BibitemShut {NoStop}%
\bibitem [{\citenamefont {Wai}\ \emph {et~al.}(1986)\citenamefont {Wai},
  \citenamefont {Menyuk}, \citenamefont {Lee},\ and\ \citenamefont
  {Chen}}]{Wai:OL:1986}%
  \BibitemOpen
  \bibfield  {author} {\bibinfo {author} {\bibfnamefont {P.~K.~A.}\
  \bibnamefont {Wai}}, \bibinfo {author} {\bibfnamefont {C.~R.}\ \bibnamefont
  {Menyuk}}, \bibinfo {author} {\bibfnamefont {Y.~C.}\ \bibnamefont {Lee}},\
  and\ \bibinfo {author} {\bibfnamefont {H.~H.}\ \bibnamefont {Chen}},\
  }\bibfield  {title} {\bibinfo {title} {Nonlinear pulse propagation in the
  neighborhood of the zero-dispersion wavelength of monomode optical fibers},\
  }\href@noop {} {\bibfield  {journal} {\bibinfo  {journal} {Opt. Lett.}\
  }\textbf {\bibinfo {volume} {11}},\ \bibinfo {pages} {464} (\bibinfo {year}
  {1986})}\BibitemShut {NoStop}%
\bibitem [{\citenamefont {Akhmediev}\ and\ \citenamefont
  {Karlsson}(1995)}]{Akhmediev:PRA:1995}%
  \BibitemOpen
  \bibfield  {author} {\bibinfo {author} {\bibfnamefont {N.}~\bibnamefont
  {Akhmediev}}\ and\ \bibinfo {author} {\bibfnamefont {M.}~\bibnamefont
  {Karlsson}},\ }\bibfield  {title} {\bibinfo {title} {Cherenkov radiation
  emitted by solitons in optical fibers},\ }\href@noop {} {\bibfield  {journal}
  {\bibinfo  {journal} {Phys. Rev. A}\ }\textbf {\bibinfo {volume} {51}},\
  \bibinfo {pages} {2602} (\bibinfo {year} {1995})}\BibitemShut {NoStop}%
\bibitem [{\citenamefont {Yulin}\ \emph {et~al.}(2004)\citenamefont {Yulin},
  \citenamefont {Skryabin},\ and\ \citenamefont {Russell}}]{Yulin:OL:2004}%
  \BibitemOpen
  \bibfield  {author} {\bibinfo {author} {\bibfnamefont {A.~V.}\ \bibnamefont
  {Yulin}}, \bibinfo {author} {\bibfnamefont {D.~V.}\ \bibnamefont
  {Skryabin}},\ and\ \bibinfo {author} {\bibfnamefont {P.~S.~J.}\ \bibnamefont
  {Russell}},\ }\bibfield  {title} {\bibinfo {title} {Four-wave mixing of
  linear waves and solitons in fibers with higher-order dispersion},\
  }\href@noop {} {\bibfield  {journal} {\bibinfo  {journal} {Opt. Lett.}\
  }\textbf {\bibinfo {volume} {29}},\ \bibinfo {pages} {2411} (\bibinfo {year}
  {2004})}\BibitemShut {NoStop}%
\bibitem [{\citenamefont {Skryabin}\ and\ \citenamefont
  {Yulin}(2005)}]{Skryabin:PRE:2005}%
  \BibitemOpen
  \bibfield  {author} {\bibinfo {author} {\bibfnamefont {D.~V.}\ \bibnamefont
  {Skryabin}}\ and\ \bibinfo {author} {\bibfnamefont {A.~V.}\ \bibnamefont
  {Yulin}},\ }\bibfield  {title} {\bibinfo {title} {Theory of generation of new
  frequencies by mixing of solitons and dispersive waves in optical fibers},\
  }\href@noop {} {\bibfield  {journal} {\bibinfo  {journal} {Phys. Rev. E}\
  }\textbf {\bibinfo {volume} {72}},\ \bibinfo {pages} {016619} (\bibinfo
  {year} {2005})}\BibitemShut {NoStop}%
\bibitem [{\citenamefont {Akhmediev}\ \emph {et~al.}(1994)\citenamefont
  {Akhmediev}, \citenamefont {Buryak},\ and\ \citenamefont
  {Karlsson}}]{Akhmediev:OC:1994}%
  \BibitemOpen
  \bibfield  {author} {\bibinfo {author} {\bibfnamefont {N.}~\bibnamefont
  {Akhmediev}}, \bibinfo {author} {\bibfnamefont {A.}~\bibnamefont {Buryak}},\
  and\ \bibinfo {author} {\bibfnamefont {M.}~\bibnamefont {Karlsson}},\
  }\bibfield  {title} {\bibinfo {title} {Radiationless optical solitons with
  oscillating tails},\ }\href@noop {} {\bibfield  {journal} {\bibinfo
  {journal} {Opt. Commun.}\ }\textbf {\bibinfo {volume} {110}},\ \bibinfo
  {pages} {540} (\bibinfo {year} {1994})}\BibitemShut {NoStop}%
\bibitem [{\citenamefont {Buryak}\ and\ \citenamefont
  {Akhmediev}(1995)}]{Buryak:PRE:1995}%
  \BibitemOpen
  \bibfield  {author} {\bibinfo {author} {\bibfnamefont {A.~V.}\ \bibnamefont
  {Buryak}}\ and\ \bibinfo {author} {\bibfnamefont {N.~N.}\ \bibnamefont
  {Akhmediev}},\ }\bibfield  {title} {\bibinfo {title} {Stability criterion for
  stationary bound states of solitons with radiationless oscillating tails},\
  }\href@noop {} {\bibfield  {journal} {\bibinfo  {journal} {Phys. Rev. E}\
  }\textbf {\bibinfo {volume} {51}},\ \bibinfo {pages} {3572} (\bibinfo {year}
  {1995})}\BibitemShut {NoStop}%
\bibitem [{\citenamefont {Pich\'e}\ \emph {et~al.}(1996)\citenamefont
  {Pich\'e}, \citenamefont {Cormier},\ and\ \citenamefont
  {Zhu}}]{Piche:OL:1996}%
  \BibitemOpen
  \bibfield  {author} {\bibinfo {author} {\bibfnamefont {M.}~\bibnamefont
  {Pich\'e}}, \bibinfo {author} {\bibfnamefont {J.-F.}\ \bibnamefont
  {Cormier}},\ and\ \bibinfo {author} {\bibfnamefont {X.}~\bibnamefont {Zhu}},\
  }\bibfield  {title} {\bibinfo {title} {{Bright optical soliton in the
  presence of fourth-order dispersion}},\ }\href@noop {} {\bibfield  {journal}
  {\bibinfo  {journal} {Opt. Lett.}\ }\textbf {\bibinfo {volume} {21}},\
  \bibinfo {pages} {845} (\bibinfo {year} {1996})}\BibitemShut {NoStop}%
\bibitem [{\citenamefont {Kruglov}(2020)}]{Kruglov:OC:2020}%
  \BibitemOpen
  \bibfield  {author} {\bibinfo {author} {\bibfnamefont {V.~I.}\ \bibnamefont
  {Kruglov}},\ }\bibfield  {title} {\bibinfo {title} {{Solitary wave and
  periodic solutions of nonlinear Schrödinger equation including higher order
  dispersions}},\ }\href@noop {} {\bibfield  {journal} {\bibinfo  {journal}
  {Opt. Commun.}\ }\textbf {\bibinfo {volume} {472}},\ \bibinfo {pages}
  {125866} (\bibinfo {year} {2020})}\BibitemShut {NoStop}%
\bibitem [{\citenamefont {Kruglov}\ and\ \citenamefont
  {Triki}(2020)}]{Kruglov:PRA:2020}%
  \BibitemOpen
  \bibfield  {author} {\bibinfo {author} {\bibfnamefont {V.~I.}\ \bibnamefont
  {Kruglov}}\ and\ \bibinfo {author} {\bibfnamefont {H.}~\bibnamefont
  {Triki}},\ }\bibfield  {title} {\bibinfo {title} {Quartic and dipole solitons
  in a highly dispersive optical waveguide with self-steepening nonlinearity
  and varying parameters},\ }\href@noop {} {\bibfield  {journal} {\bibinfo
  {journal} {Phys. Rev. A}\ }\textbf {\bibinfo {volume} {102}},\ \bibinfo
  {pages} {043509} (\bibinfo {year} {2020})}\BibitemShut {NoStop}%
\bibitem [{\citenamefont {Triki}\ and\ \citenamefont
  {Kruglov}(2020)}]{Triki:PRE:2020}%
  \BibitemOpen
  \bibfield  {author} {\bibinfo {author} {\bibfnamefont {H.}~\bibnamefont
  {Triki}}\ and\ \bibinfo {author} {\bibfnamefont {V.~I.}\ \bibnamefont
  {Kruglov}},\ }\bibfield  {title} {\bibinfo {title} {Propagation of dipole
  solitons in inhomogeneous highly dispersive optical-fiber media},\
  }\href@noop {} {\bibfield  {journal} {\bibinfo  {journal} {Phys. Rev. E}\
  }\textbf {\bibinfo {volume} {101}},\ \bibinfo {pages} {042220} (\bibinfo
  {year} {2020})}\BibitemShut {NoStop}%
\bibitem [{\citenamefont {Kruglov}\ and\ \citenamefont
  {Triki}(2023)}]{Kruglov:CSF:2023}%
  \BibitemOpen
  \bibfield  {author} {\bibinfo {author} {\bibfnamefont {V.~I.}\ \bibnamefont
  {Kruglov}}\ and\ \bibinfo {author} {\bibfnamefont {H.}~\bibnamefont
  {Triki}},\ }\bibfield  {title} {\bibinfo {title} {Propagation of coupled
  quartic and dipole multi-solitons in optical fibers medium with higher-order
  dispersions},\ }\href@noop {} {\bibfield  {journal} {\bibinfo  {journal}
  {Chaos, Solitons \& Fractals}\ }\textbf {\bibinfo {volume} {172}},\ \bibinfo
  {pages} {113526} (\bibinfo {year} {2023})}\BibitemShut {NoStop}%
\bibitem [{\citenamefont {Kruglov}\ and\ \citenamefont
  {Triki}(2022)}]{Kruglov:CSF:2022}%
  \BibitemOpen
  \bibfield  {author} {\bibinfo {author} {\bibfnamefont {V.~I.}\ \bibnamefont
  {Kruglov}}\ and\ \bibinfo {author} {\bibfnamefont {H.}~\bibnamefont
  {Triki}},\ }\bibfield  {title} {\bibinfo {title} {Propagation of periodic and
  solitary waves in a highly dispersive cubic–quintic medium with
  self-frequency shift and self-steepening nonlinearity},\ }\href@noop {}
  {\bibfield  {journal} {\bibinfo  {journal} {Chaos, Solitons \& Fractals}\
  }\textbf {\bibinfo {volume} {164}},\ \bibinfo {pages} {112704} (\bibinfo
  {year} {2022})}\BibitemShut {NoStop}%
\bibitem [{\citenamefont {Tam}\ \emph {et~al.}(2019)\citenamefont {Tam},
  \citenamefont {Alexander}, \citenamefont {Blanco-Redondo},\ and\
  \citenamefont {de~Sterke}}]{Tam:OL:2019}%
  \BibitemOpen
  \bibfield  {author} {\bibinfo {author} {\bibfnamefont {K.~K.~K.}\
  \bibnamefont {Tam}}, \bibinfo {author} {\bibfnamefont {T.~J.}\ \bibnamefont
  {Alexander}}, \bibinfo {author} {\bibfnamefont {A.}~\bibnamefont
  {Blanco-Redondo}},\ and\ \bibinfo {author} {\bibfnamefont {C.~M.}\
  \bibnamefont {de~Sterke}},\ }\bibfield  {title} {\bibinfo {title} {Stationary
  and dynamical properties of pure-quartic solitons},\ }\href@noop {}
  {\bibfield  {journal} {\bibinfo  {journal} {Opt. Lett.}\ }\textbf {\bibinfo
  {volume} {44}},\ \bibinfo {pages} {3306} (\bibinfo {year}
  {2019})}\BibitemShut {NoStop}%
\bibitem [{\citenamefont {Ablowitz}\ and\ \citenamefont
  {Musslimani}(2005)}]{Ablowitz:OL:2005}%
  \BibitemOpen
  \bibfield  {author} {\bibinfo {author} {\bibfnamefont {M.~J.}\ \bibnamefont
  {Ablowitz}}\ and\ \bibinfo {author} {\bibfnamefont {Z.~H.}\ \bibnamefont
  {Musslimani}},\ }\bibfield  {title} {\bibinfo {title} {Spectral
  renormalization method for computing self-localized solutions to nonlinear
  systems},\ }\href@noop {} {\bibfield  {journal} {\bibinfo  {journal} {Opt.
  Lett.}\ }\textbf {\bibinfo {volume} {30}},\ \bibinfo {pages} {2140} (\bibinfo
  {year} {2005})}\BibitemShut {NoStop}%
\bibitem [{\citenamefont {Petviashvili}(1976)}]{Petviashvili:SJPP:1976}%
  \BibitemOpen
  \bibfield  {author} {\bibinfo {author} {\bibfnamefont {V.~I.}\ \bibnamefont
  {Petviashvili}},\ }\bibfield  {title} {\bibinfo {title} {Equation for an
  extraordinary soliton},\ }\href@noop {} {\bibfield  {journal} {\bibinfo
  {journal} {Sov. J. Plasma Phys.}\ }\textbf {\bibinfo {volume} {2}},\ \bibinfo
  {pages} {257} (\bibinfo {year} {1976})}\BibitemShut {NoStop}%
\bibitem [{\citenamefont {Pelinovsky}\ and\ \citenamefont
  {Kivshar}(2000)}]{Pelinovsky:PRE:2000}%
  \BibitemOpen
  \bibfield  {author} {\bibinfo {author} {\bibfnamefont {D.}~\bibnamefont
  {Pelinovsky}}\ and\ \bibinfo {author} {\bibfnamefont {Y.}~\bibnamefont
  {Kivshar}},\ }\bibfield  {title} {\bibinfo {title} {Stability criterion for
  multicomponent solitary waves},\ }\href@noop {} {\bibfield  {journal}
  {\bibinfo  {journal} {Phys. Rev. E}\ }\textbf {\bibinfo {volume} {62}},\
  \bibinfo {pages} {8668} (\bibinfo {year} {2000})}\BibitemShut {NoStop}%
\bibitem [{\citenamefont {Musslimani}\ and\ \citenamefont
  {Yang}(2004)}]{Musslimani:JOSAB:2004}%
  \BibitemOpen
  \bibfield  {author} {\bibinfo {author} {\bibfnamefont {Z.}~\bibnamefont
  {Musslimani}}\ and\ \bibinfo {author} {\bibfnamefont {J.}~\bibnamefont
  {Yang}},\ }\bibfield  {title} {\bibinfo {title} {Self-trapping of light in a
  two-dimensional photonic lattice},\ }\href@noop {} {\bibfield  {journal}
  {\bibinfo  {journal} {J. Opt. Soc. Am. B}\ }\textbf {\bibinfo {volume}
  {21}},\ \bibinfo {pages} {973} (\bibinfo {year} {2004})}\BibitemShut
  {NoStop}%
\bibitem [{\citenamefont {Fibich}\ \emph {et~al.}(2006)\citenamefont {Fibich},
  \citenamefont {Sivan},\ and\ \citenamefont {Weinstein}}]{Fibich:PD:2006}%
  \BibitemOpen
  \bibfield  {author} {\bibinfo {author} {\bibfnamefont {G.}~\bibnamefont
  {Fibich}}, \bibinfo {author} {\bibfnamefont {Y.}~\bibnamefont {Sivan}},\ and\
  \bibinfo {author} {\bibfnamefont {M.~I.}\ \bibnamefont {Weinstein}},\
  }\bibfield  {title} {\bibinfo {title} {{Bound states of nonlinear
  Schrödinger equations with a periodic nonlinear microstructure}},\
  }\href@noop {} {\bibfield  {journal} {\bibinfo  {journal} {Physica D}\
  }\textbf {\bibinfo {volume} {217}},\ \bibinfo {pages} {31} (\bibinfo {year}
  {2006})}\BibitemShut {NoStop}%
\bibitem [{\citenamefont {Amiranashvili}\ \emph {et~al.}(2013)\citenamefont
  {Amiranashvili}, \citenamefont {Bandelow},\ and\ \citenamefont
  {Akhmediev}}]{Amiranashvili:PRA:2013}%
  \BibitemOpen
  \bibfield  {author} {\bibinfo {author} {\bibfnamefont {S.}~\bibnamefont
  {Amiranashvili}}, \bibinfo {author} {\bibfnamefont {U.}~\bibnamefont
  {Bandelow}},\ and\ \bibinfo {author} {\bibfnamefont {N.}~\bibnamefont
  {Akhmediev}},\ }\bibfield  {title} {\bibinfo {title} {Few-cycle optical
  solitary waves in nonlinear dispersive media},\ }\href@noop {} {\bibfield
  {journal} {\bibinfo  {journal} {Phys. Rev. A}\ }\textbf {\bibinfo {volume}
  {87}},\ \bibinfo {pages} {013805} (\bibinfo {year} {2013})}\BibitemShut
  {NoStop}%
\bibitem [{\citenamefont {Heidt}(2009)}]{Heidt:JLT:2009}%
  \BibitemOpen
  \bibfield  {author} {\bibinfo {author} {\bibfnamefont {A.~M.}\ \bibnamefont
  {Heidt}},\ }\bibfield  {title} {\bibinfo {title} {Efficient adaptive step
  size method for the simulation of supercontinuum generation in optical
  fibers},\ }\href@noop {} {\bibfield  {journal} {\bibinfo  {journal} {IEEE J.
  Lightwave Tech.}\ }\textbf {\bibinfo {volume} {27}},\ \bibinfo {pages} {3984}
  (\bibinfo {year} {2009})}\BibitemShut {NoStop}%
\bibitem [{\citenamefont {Melchert}\ and\ \citenamefont
  {Demircan}(2022)}]{Melchert:CPC:2022}%
  \BibitemOpen
  \bibfield  {author} {\bibinfo {author} {\bibfnamefont {O.}~\bibnamefont
  {Melchert}}\ and\ \bibinfo {author} {\bibfnamefont {A.}~\bibnamefont
  {Demircan}},\ }\bibfield  {title} {\bibinfo {title} {{py-fmas: A python
  package for ultrashort optical pulse propagation in terms of forward models
  for the analytic signal}},\ }\href@noop {} {\bibfield  {journal} {\bibinfo
  {journal} {Computer Physics Communications}\ }\textbf {\bibinfo {volume}
  {273}},\ \bibinfo {pages} {108257} (\bibinfo {year} {2022})}\BibitemShut
  {NoStop}%
\bibitem [{\citenamefont {Hult}(2007)}]{Hult:JLT:2007}%
  \BibitemOpen
  \bibfield  {author} {\bibinfo {author} {\bibfnamefont {J.}~\bibnamefont
  {Hult}},\ }\bibfield  {title} {\bibinfo {title} {{A Fourth-Order
  Runge–Kutta in the Interaction Picture Method for Simulating Supercontinuum
  Generation in Optical Fibers}},\ }\href@noop {} {\bibfield  {journal}
  {\bibinfo  {journal} {IEEE J. Lightwave Tech.}\ }\textbf {\bibinfo {volume}
  {25}},\ \bibinfo {pages} {3770} (\bibinfo {year} {2007})}\BibitemShut
  {NoStop}%
\bibitem [{\citenamefont {Melchert}\ \emph {et~al.}(2020)\citenamefont
  {Melchert}, \citenamefont {Yulin},\ and\ \citenamefont
  {Demircan}}]{Melchert:OL:2020}%
  \BibitemOpen
  \bibfield  {author} {\bibinfo {author} {\bibfnamefont {O.}~\bibnamefont
  {Melchert}}, \bibinfo {author} {\bibfnamefont {A.}~\bibnamefont {Yulin}},\
  and\ \bibinfo {author} {\bibfnamefont {A.}~\bibnamefont {Demircan}},\
  }\bibfield  {title} {\bibinfo {title} {{Dynamics of localized dissipative
  structures in a generalized Lugiato–Lefever model with negative quartic
  group-velocity dispersion}},\ }\href@noop {} {\bibfield  {journal} {\bibinfo
  {journal} {Opt. Lett.}\ }\textbf {\bibinfo {volume} {45}},\ \bibinfo {pages}
  {2764} (\bibinfo {year} {2020})}\BibitemShut {NoStop}%
\bibitem [{\citenamefont {Parra-Rivas}\ \emph {et~al.}(2014)\citenamefont
  {Parra-Rivas}, \citenamefont {Gomila}, \citenamefont {Mat\'{\i}as},
  \citenamefont {Coen},\ and\ \citenamefont {Gelens}}]{Rivas:PRA:2014}%
  \BibitemOpen
  \bibfield  {author} {\bibinfo {author} {\bibfnamefont {P.}~\bibnamefont
  {Parra-Rivas}}, \bibinfo {author} {\bibfnamefont {D.}~\bibnamefont {Gomila}},
  \bibinfo {author} {\bibfnamefont {M.~A.}\ \bibnamefont {Mat\'{\i}as}},
  \bibinfo {author} {\bibfnamefont {S.}~\bibnamefont {Coen}},\ and\ \bibinfo
  {author} {\bibfnamefont {L.}~\bibnamefont {Gelens}},\ }\bibfield  {title}
  {\bibinfo {title} {{Dynamics of localized and patterned structures in the
  Lugiato-Lefever equation determine the stability and shape of optical
  frequency combs}},\ }\href@noop {} {\bibfield  {journal} {\bibinfo  {journal}
  {Phys. Rev. A}\ }\textbf {\bibinfo {volume} {89}},\ \bibinfo {pages} {043813}
  (\bibinfo {year} {2014})}\BibitemShut {NoStop}%
\bibitem [{\citenamefont {Anderson}\ and\ \citenamefont
  {Lisak}(1986)}]{Anderson:PS:1986}%
  \BibitemOpen
  \bibfield  {author} {\bibinfo {author} {\bibfnamefont {D.}~\bibnamefont
  {Anderson}}\ and\ \bibinfo {author} {\bibfnamefont {M.}~\bibnamefont
  {Lisak}},\ }\bibfield  {title} {\bibinfo {title} {{Variational Approach to
  Incoherent Two-soliton Interactions}},\ }\href@noop {} {\bibfield  {journal}
  {\bibinfo  {journal} {Physica Scripta}\ }\textbf {\bibinfo {volume} {33}},\
  \bibinfo {pages} {193} (\bibinfo {year} {1986})}\BibitemShut {NoStop}%
\bibitem [{\citenamefont {Karpman}\ and\ \citenamefont
  {Solov'ev}(1981)}]{Karpman:PD:1981}%
  \BibitemOpen
  \bibfield  {author} {\bibinfo {author} {\bibfnamefont {V.~I.}\ \bibnamefont
  {Karpman}}\ and\ \bibinfo {author} {\bibfnamefont {V.~V.}\ \bibnamefont
  {Solov'ev}},\ }\bibfield  {title} {\bibinfo {title} {{A perturbational
  approach to the two-soliton systems}},\ }\href@noop {} {\bibfield  {journal}
  {\bibinfo  {journal} {Physica D}\ }\textbf {\bibinfo {volume} {3}},\ \bibinfo
  {pages} {487} (\bibinfo {year} {1981})}\BibitemShut {NoStop}%
\bibitem [{\citenamefont {Melchert}\ \emph {et~al.}(2021)\citenamefont
  {Melchert}, \citenamefont {Willms}, \citenamefont {Morgner}, \citenamefont
  {Babushkin},\ and\ \citenamefont {Demircan}}]{Melchert:SR:2021}%
  \BibitemOpen
  \bibfield  {author} {\bibinfo {author} {\bibfnamefont {O.}~\bibnamefont
  {Melchert}}, \bibinfo {author} {\bibfnamefont {S.}~\bibnamefont {Willms}},
  \bibinfo {author} {\bibfnamefont {U.}~\bibnamefont {Morgner}}, \bibinfo
  {author} {\bibfnamefont {I.}~\bibnamefont {Babushkin}},\ and\ \bibinfo
  {author} {\bibfnamefont {A.}~\bibnamefont {Demircan}},\ }\bibfield  {title}
  {\bibinfo {title} {Crossover from two‐frequency pulse compounds to escaping
  solitons},\ }\href@noop {} {\bibfield  {journal} {\bibinfo  {journal}
  {Scientific Reports}\ }\textbf {\bibinfo {volume} {11}},\ \bibinfo {pages}
  {11190} (\bibinfo {year} {2021})}\BibitemShut {NoStop}%
\bibitem [{\citenamefont {Anderson}\ \emph {et~al.}(1992)\citenamefont
  {Anderson}, \citenamefont {H\"o\"ok}, \citenamefont {Lisak}, \citenamefont
  {Serkin},\ and\ \citenamefont {Afanasjev}}]{Anderson:EL:1992}%
  \BibitemOpen
  \bibfield  {author} {\bibinfo {author} {\bibfnamefont {D.}~\bibnamefont
  {Anderson}}, \bibinfo {author} {\bibfnamefont {A.}~\bibnamefont {H\"o\"ok}},
  \bibinfo {author} {\bibfnamefont {M.}~\bibnamefont {Lisak}}, \bibinfo
  {author} {\bibfnamefont {V.~N.}\ \bibnamefont {Serkin}},\ and\ \bibinfo
  {author} {\bibfnamefont {V.~V.}\ \bibnamefont {Afanasjev}},\ }\bibfield
  {title} {\bibinfo {title} {{Soliton cross-trapping: new method for bright
  soliton transmission at normal group velocity dispersion}},\ }\href@noop {}
  {\bibfield  {journal} {\bibinfo  {journal} {Electron. Lett.}\ }\textbf
  {\bibinfo {volume} {28}},\ \bibinfo {pages} {1797} (\bibinfo {year}
  {1992})}\BibitemShut {NoStop}%
\bibitem [{\citenamefont {Driben}\ \emph {et~al.}(2013)\citenamefont {Driben},
  \citenamefont {Yulin}, \citenamefont {Efimov},\ and\ \citenamefont
  {Malomed}}]{Driben:OE:2013}%
  \BibitemOpen
  \bibfield  {author} {\bibinfo {author} {\bibfnamefont {R.}~\bibnamefont
  {Driben}}, \bibinfo {author} {\bibfnamefont {A.~V.}\ \bibnamefont {Yulin}},
  \bibinfo {author} {\bibfnamefont {A.}~\bibnamefont {Efimov}},\ and\ \bibinfo
  {author} {\bibfnamefont {B.~A.}\ \bibnamefont {Malomed}},\ }\bibfield
  {title} {\bibinfo {title} {Trapping of light in solitonic cavities and its
  role in the supercontinuum generation},\ }\href@noop {} {\bibfield  {journal}
  {\bibinfo  {journal} {Opt. Express}\ }\textbf {\bibinfo {volume} {21}},\
  \bibinfo {pages} {19091} (\bibinfo {year} {2013})}\BibitemShut {NoStop}%
\bibitem [{\citenamefont {Wang}\ \emph {et~al.}(2015)\citenamefont {Wang},
  \citenamefont {Mussot}, \citenamefont {Conforti}, \citenamefont {Zeng},\ and\
  \citenamefont {Kudlinski}}]{Wang:OL:2015}%
  \BibitemOpen
  \bibfield  {author} {\bibinfo {author} {\bibfnamefont {S.~F.}\ \bibnamefont
  {Wang}}, \bibinfo {author} {\bibfnamefont {A.}~\bibnamefont {Mussot}},
  \bibinfo {author} {\bibfnamefont {M.}~\bibnamefont {Conforti}}, \bibinfo
  {author} {\bibfnamefont {X.~L.}\ \bibnamefont {Zeng}},\ and\ \bibinfo
  {author} {\bibfnamefont {A.}~\bibnamefont {Kudlinski}},\ }\bibfield  {title}
  {\bibinfo {title} {Bouncing of a dispersive wave in a solitonic cage},\
  }\href@noop {} {\bibfield  {journal} {\bibinfo  {journal} {Opt. Lett.}\
  }\textbf {\bibinfo {volume} {40}},\ \bibinfo {pages} {3320} (\bibinfo {year}
  {2015})}\BibitemShut {NoStop}%
\end{thebibliography}%
\end{document}